\documentclass[a4paper,11pt]{article}
\pdfoutput=1 

\usepackage{jcappub} 

\usepackage{graphicx}
\usepackage{amssymb}
\usepackage{amsmath}
\usepackage{bm}
\usepackage{multirow}
\usepackage{cancel}
\usepackage{subfigure}
\usepackage{color}
\usepackage[version=3]{mhchem}
\usepackage{adjustbox}
\usepackage{ulem}

\newcommand{\bea}{\begin{eqnarray}}
\newcommand{\eea}{\end{eqnarray}}

\title{
Oscillating cosmic evolution and 
constraints on big bang nucleosynthesis
in the extended Starobinsky model
}

\author[a]{Jubin Park}
\author[a]{Chae-min Yun}
\author[a]{Myung-Ki Cheoun}
\author[b,1]{Dukjae Jang \note{Corresponding author.}}

\affiliation[a]{Department of Physics and Origin of Matter and Evolution of Galaxies (OMEG) Institute, Soongsil University, Seoul 06978, Republic of Korea}
\affiliation[b]{Center for Relativistic Laser Science, Institute for Basic Science (IBS), Gwangju 61005, Republic of Korea}

\emailAdd{honolov@ssu.ac.kr}
\emailAdd{c.yun@ssu.ac.kr}
\emailAdd{cheoun@ssu.ac.kr}
\emailAdd{djjang2@ibs.re.kr}

\abstract{We investigate the cosmic evolutions in the extended Starobinsky model (eSM) obtained by adding one $R^{ab}R_{ab}$ term to the Starobinsky model. We discuss the possibility of various cosmic evolutions with a special focus on the radiation-dominated era (RDE). Using simple assumptions, a second-order non-linear differential equation describing the various cosmic evolutions in the eSM is introduced. By solving this non-linear equation numerically, we show that the various cosmic evolutions, such as the standard cosmic evolution ($a \propto t^{1/2}$) and a unique oscillating cosmic evolution, are feasible due to the effects of higher-order terms introduced beyond Einstein's gravity. Furthermore, we consider big bang nucleosynthesis (BBN), which is the most important observational result in the RDE, to constrain the free parameters of the eSM. The primordial abundances of the light elements, such as $^{4}$He, D, $^{3}$He, $^{7}$Li, and $^{6}$Li by the cosmic evolutions are compared with the most recent observational data. It turns out that most non-standard cosmic evolutions can not easily satisfy these BBN constraints, but a free parameter of the viable models with the oscillating cosmic evolution is shown to have an upper limit by the constraints. In particular, we find that the free parameter is most sensitive to deuterium and $^4$He abundances, which are being precisely measured among other elements. Therefore, more accurate measurements in the near future may enable us to distinguish the eSM from the standard model as well as other models.}


\begin{document}
\maketitle
\flushbottom

\section{INTRODUCTION}
The current standard cosmological model is based on Einstein's theory of gravity with the Friedmann-Lema\^itre-Robertson-Walker (FLRW) metric for the homogeneous and isotropic universe.
All known standard model particles, as well as dark matter and dark energy, are carefully contained in this model. 
More importantly, three obvious pieces of historical evidence, big bang nucleosynthesis (BBN), the cosmic microwave background (CMB) and the expansion of the universe strongly support the standard cosmological model.
Among them, BBN provides the earliest information that occurred between 1 second and 300 seconds after the birth of the universe. 
Compared to the data provided by CMB photons in the matter-dominated era (MDE) after 380,000 years, the BBN observations are the only observable ones in the radiation-dominated era (RDE). 
Therefore, observation of light element abundances in old astronomical bodies is one of the most important premises for the cosmological model to be verified.

So far, it is known that theoretical predictions of the abundance of some light elements are in good agreement with the observed data.
In recent years, the precision of light elements abundances such as  $^4$He and deuterium (D) has improved at a faster rate than before, reaching the percentage level.
As a result, the predictions and observations of light elements in the RDE not only validate the standard cosmic nucleosynthesis, but also enable us to permit constraints of various non-standard nucleosynthesis models and discriminate the extended models from the standard cosmological model.

Recently, the modified gravity (MG) models have received a lot of attention \cite{PRD08Weinberg, 05 H and N,  Nojiri:2010wj, Nojiri:2017ncd}, 
because they can provide solutions not only to the problems related to the observation of cosmic microwave background radiation, but also to the horizon problem and flatness problem\,\cite{Starobinsky:1980te}.
Interestingly, the MG models also provide several changes in the early universe.
They may include the ingenious ways to cause new cosmic evolutions beyond the standard cosmic evolution in the very early, radiation-dominated, matter-dominated times, and even today.
If the cosmic evolutions of the MG models in the RDE can be (significantly) different from that of the standard cosmological model, the MG effect will (noticeably) affect BBN and the predicted abundances of light elements will be different from those of the standard BBN (SBBN) model.

Since all the physics involved in the SBBN is already well established, the deviations from the BBN observations may provide hints or constraints on new physics and/or astrophysics in the early universe. 
In particular, the formation of primordial abundances is sensitive to change in the cosmic expansion rate, so gravity models beyond Einstein's theory with the FLRW universe have been explored to evaluate viability of those models in the BBN epoch.
For representative examples, the implication of the $f(R)\propto R^n$ type MG models\,\cite{Clifton:2005aj,Lambiase:2006dq,Kang:2008zi,Kusakabe:2015yaa} and the modified Gauss-Bonnet gravity\,\cite{Kusakabe:2015ida} on BBN has been studied, which lead to constraints on the free parameters in the given MG models such as index $n$ based on the BBN observations.
Moreover, for the braneworld universe proposed to solve the hierarchy problem among the fundamental forces \cite{Randall:1999vf, Randall:1999ee}, the effects of modified cosmological evolution \cite{Binetruy:1999hy, Langlois:2000ph, Langlois:2000ia} on BBN were investigated \cite{Sasankan:2016ixg, Jang:2016rpi}.

In this study, we investigate the extended Starobinsky model (eSM) which added one $R^{ab}R_{ab}$ term to the Starobinsky model.
We discuss the possibility of various cosmic evolutions by the eSM in the RDE. 
Mainly we focus on the non-standard cosmic evolutions, such as the \textbf{\textit{oscillating cosmic evolutions}} beyond the standard cosmic one, which typically has a scale factor of $a(t) \propto t^{1/2}$. 
Afterwards, the BBN observations are discussed to constrain the free parameters in the eSM.
For this purpose, the recently released \texttt{Mathematica} BBN code \texttt{PRIMAT}\,\cite{Pitrou:2018cgg,Pitrou:2020etk} is used because of its improved precision for the primordial abundances of D and $^4$He.
The neutron lifetime and the numerous corrections to the weak interaction processes are incorporated into the code to achieve this percent-level precision. 
More specifically, the radiative, zero-temperature, finite nucleon mass, finite temperature radiative, weak-magnetism corrections, and QED plasma effects are considered in a self-consistent way\,\cite{Pitrou:2018cgg}.
In particular, \texttt{PRIMAT} has computed the precision of the primordial abundances of ${\rm Y_p}$ and D/H with a relative uncertainty of approximately 0.01 (1\%), enabling us to precisely constrain the free parameters in the eSM.
The only free parameter $\mathcal{C}$ given as the sum of the coefficients of $R^{ab}R_{ab}$ and $R^2$ is then constrained.

Here, we briefly introduce and summarize the important points of this study:
\begin{itemize}
\item
In the RDE, the non-standard cosmic evolutions and BBN constraints are studied in the eSM.

\item 
One \textit{second-order} \textbf{\textit{non-linear}} differential equation describing the cosmic evolutions in the eSM is introduced under several assumptions. By solving this non-linear equation numerically, both solutions of the standard and interesting non-standard cosmic evolution are obtained.

\item
We discover three distinct types of interesting non-standard cosmic evolution solutions in the RDE.
The first type represents an exponential expansion of the scale factor (similar to the role of the cosmological constant term), and the second one describes the stationary universe in which the size of the universe does not change after a specific time.
Finally, the third one represents the interesting \textbf{\textit{oscillating cosmic evolutions}} around the standard cosmic evolution.

\item
Except for the third type, it turns out that the first and second types can not comply with 
the current BBN observations.

\item
Among the abundances of the light elements, the abundances of D and $^4$He are most sensitive to the $\mathcal{C}$ parameter in the eSM.

\item
In the viable eSM permitting \textbf{\textit{oscillating cosmic evolution}}, the constraints of the parameter $\mathcal{C}$ are obtained. 
Interestingly, the negative region of $\mathcal{C}$ is excluded and the allowed region of $\mathcal{C}$ should be very narrow; $0 < \mathcal{C} < 1.3 \times 10^{-5}$ ${\rm s^2}$, where ${\rm s}$ represents the unit of time, seconds.

\item
In the near future, it may be possible to distinguish the eSM and other extended models from the SBBN model with more accurately measured abundances and precise predictions.\footnote{
CMB anisotropies or large scale structures or other important observables 
may be much better for distinguishing the extended models 
if the evolution of the universe deviates from the standard evolution and is maintained until late times. However, in the case of the eSM model, higher-order terms can quickly disappear during the transition from the radiation-dominant period to the material-dominant period \cite{Yun:2022xce}.}
\end{itemize}

This paper is organized as follows. 
In Sec.\,II, 
the aforementioned eSM is introduced, and 
one \textit{second-order} \textbf{\textit{non-linear}} differential equation 
describing the cosmic evolutions is reviewed. 
In Sec.\,III,
the standard and non-standard cosmic evolutions in the eSM are discussed.
In Sec.\,IV,
the BBN calculation method in the eSM with the \textbf{\textit{oscillating cosmic evolutions}}
is explained.
In Sec.\,V,
the observational abundances of light elements 
are briefly described.
In Sec.\,VI, 
the results of the SBBN and the three benchmark models are presented and 
the constraint conditions of the model parameter $\mathcal{C}$ are extracted.
In Sec.\,VII,
we provide a summary and conclusions of this study.

\section{Extended Starobinsky gravity model}
In this eSM, the cosmological equations are derived from the following action
\cite{97 N and H, 99 N and H, 05Mukhanov, 10 Sotiriou and Faraoni, Buchbinder, 82 birrell and davis, 84 Whitt, 08 Capozziello and De Felice}
\bea
&& S    =  \int d^4x \sqrt{-g} \Big[ {1 \over 16\pi G } 
\big( R + \mathcal{A} R^2 + \mathcal{B} R^{ab} R_{ab}  \big)  + L_m \Big]  \label{action} ,
\eea
where $R$ is the Einstein-Hilbert action representing Einstein's gravity, 
and terms $\mathcal{A} R^2$ (hereafter referred to as $\mathcal{A}$ term) and 
$\mathcal{B} R^{ab} R_{ab}$ (referred to as $\mathcal{B}$ term) are the additional 
higher order terms. 
The ${L_m}$ is the matter part Lagrangian for radiation and matter in the early 
universe\footnote{
$ \delta \big(\sqrt{-g}L_m  \big)  \equiv {1 \over 2} \sqrt{-g} T^{ab} \delta g_{ab}$.}.
We follow the Hawking-Ellis convention \cite{Hawking and Ellis}
and set 
$\hbar=k_B=c=1$.
As a specific example of the $f(R)$ gravity theory, 
Starobinsky gravity model where $\mathcal{B}=L_m=0$ is one of the preferred 
cosmological models 
\,\cite{Starobinsky:1980te}
\footnote{
This is due to the results of Planck satellite experiment, 
which precisely measures the anisotropy of CMB radiation\,\cite{Planck2018}.}.
It is worth noting that the additional $\mathcal{B} R^{ab} R_{ab}$ term 
is neither $f(R)$ gravity nor conformally symmetric\,\cite{97 N and H}, but 
has effects similar to those of $\mathcal{A} R^2$ 
in cosmic evolution.

The variation in action (\ref{action}) \cite{72Weinberg, 83Barth} results in the following 
gravitational field equations (GFE) \cite{97 N and H} 
\bea
&&  R_{ab} - {1 \over 2} g_{ab} R - 8\pi G ( T^\mathcal{A}_{ab}   + 
T^\mathcal{B}_{ab} )  = 8\pi G T_{ab}     
          \label{GFE}   	,
\eea 
where $T^{\mathcal{A}}_{ab}$ and $T^{\mathcal{B}}_{ab}$ are given as
\bea	
 &&  T^\mathcal{A}_{ab} \equiv {\mathcal{A} \over {8\pi G} } \Big( {1 \over 2}  R^{2} g_{ab} - 2R R_{ab} - 2 g_{ab} \Box R + 2 R_{;ab}   \Big)    ,
\label{EM tensor A}
\\ && T^\mathcal{B}_{ab} \equiv  
{\mathcal{B} \over {8\pi G}} \Big(  
            {1 \over 2 } R^{cd} R_{cd} g_{ab} - g_{ab}  {R^{cd}}_{;cd}  
		      + 2  {R}^{c}_{ \phantom{c} (a ;b)c} -  \Box R_{ab} - 2  {R_a}^c R_{bc}   \Big)
\nonumber \\ && 
\phantom{T^\mathcal{B}_{ab}}
= {\mathcal{B} \over {8\pi G}} \Big( {1 \over 2}  R^{cd} R_{cd} g_{ab} + R_{;ab} - 2 R^{cd} R_{acbd} - {1 \over 2} g_{ab} \Box R - \Box R_{ab}  \Big)  {}
                \label{EM tensor B}                    . \eea\
Here, the d'Alembert operator is denoted by a box (e.g. $\Box R \equiv g^{cd} R_{;cd}$), and 
the semicolons and subscripts with parentheses 
stand for the covariant derivatives and the symmetrization of 
a tensor\,\footnote{
For example, $t_{(ab)} \equiv {1 \over 2} (t_{ab} + t_{ba})$.
}, respectively.
In Eq.\,(\ref{EM tensor B}), the Bianchi identity $ R^b_{a;b} = {1 \over 2} R_{;a} $ \cite{Hawking and Ellis, 72Weinberg} and the commutation relation for covariant derivatives of a tensor 
$ T_{a \phantom{b} ;nm}^{\phantom{a}b} - T_{a \phantom{b} ;mn}^{\phantom{a}b} 
 = T_r^{\phantom{r} b} R_{a \phantom{r} mn}^{\phantom{a} r} + T_a^{\phantom{a} r} R^b_{\phantom{b} rmn}  $ \cite{83Barth}
are applied in order to conveniently read off each component of the energy-momentum tensor by reproducing the previous calculations from the Appendices in Ref.\,\cite{97 N and H}.

Adopting the spatially flat FLRW metric 
for a homogeneous and isotropic universe, the infinitesimal interval $(ds)$ is given by
\bea 
 && ds^2 = -dt^2 + a(t)^2  \delta_{\alpha \beta} dx^{\alpha} dx^{\beta}   \label{metric} ,
\eea
where the $a(t)$ is the cosmic scale factor that presents the relative expansion of the 
universe.
The energy-momentum tensor consisting of energy density $(\rho)$ and pressure $(p)$ 
is written component by component as follows
\bea
T^0_0 = - \rho , 
\quad T^0_{\alpha} = 0,
 \quad T^\alpha_\beta = p \delta^\alpha_\beta   
\label{EM tensor}  .
\eea
The time and space components of the GFE in Eq.\,(\ref{GFE}) become\,\cite{97 N and H}
\bea
 && 8\pi G \rho = 3 H^2 + 6 (3\mathcal{A}+\mathcal{B}) 
 \big( 2 \ddot H H - \dot H ^2  +  6 \dot H H^2 \big)  
\label{Friedmann eq} ,
\\ && 8\pi G p  = -  ( 2\dot H + 3 H^2) 
  - 2 (3\mathcal{A}+\mathcal{B}) 
  \bigg( 2 { {d^3  H} \over {dt^3}} + 12 \ddot H H + 9 \dot H ^2 + 18 \dot H H^2 \bigg)   
\label{accel eq}
\eea
where the dot over the dependent variable represents the derivative with respect to time,
and $H \equiv \dot a / a$ represents the cosmic expansion rate.
Interestingly, in the above two equations, since the coefficients 
$\mathcal{A}$ and $\mathcal{B}$ are grouped together in parentheses, 
they can be expressed with only one free parameter $\mathcal{C} 
\equiv (3\mathcal{A}+\mathcal{B})$.
The consistency can be checked by substituting these two modified Friedmann equations in
Eqs.\,(\ref{Friedmann eq}) and (\ref{accel eq})
into the supplementary conservation equation
\bea
\dot \rho + 3 H (\rho + p ) = 0
\label{conservation eq} .
\eea
Eq.\,(\ref{conservation eq}) comes from $T^a_{b;a} = 0$, 
which is implied by the divergence-free property of the left hand side of the GFE in Eq.\,(\ref{GFE}) (For more detail, see Appendix C).
The time derivative of Eq.\,(\ref{Friedmann eq}) becomes
\bea 
 8\pi G \dot \rho 
  = 6 \dot H H 
+ 12 \,\mathcal{C} H \Big( { {d^3  H} \over {dt^3}}  + 3 \ddot{H} H + 6 \dot{H}^2  \Big)
    \label{Dot-Friedmann eq}.
\eea
We assume that $\rho = \rho_r$ during the RDE with the equation of state (EOS) for radiation, $p_r = \rho_r/3$, 
where $\rho_r$ and $p_r$ are, respectively, the energy density and pressure of only relativistic 
particles \,\footnote{
In principle, the (total) energy density $\rho$ should contain all known types of energy components, i.e.
$\rho = \rho_r + \rho_b + \rho_{cdm} + \rho_\Lambda$.
Here, $\rho_r$ includes the (relativistic) contributions of about three active neutrinos $(\rho_\nu)$, photons $(\rho_\gamma)$, and 
the relativistic plasma components $(\rho_{pl})$. 
$\rho_b\ (\rho_{cdm})$ represents the non-relativistic energy contribution of baryons (cold species).
The $\rho_\Lambda$ represents the energy density of the cosmological constant.
However, the components except $\rho_r$ are negligibly small compared to $\rho_r$ 
during the RDE in which BBN occurs.
For example, in the BBN period of the SBBN model, $\rho_b$ is approximately $10^{-4 \sim -6}$ smaller 
than $\rho_r$, and $\rho_\Lambda$ is much smaller by $\sim 10^{-30}$.
Therefore, the assumption of $\rho=\rho_r$ is reasonable. On the other hand, the assumed EOS of $p_r = \rho_r/3$ could somewhat differ from the EOS in precise BBN calculation. This is because electrons and positrons change from relativistic to non-relativistic species during the BBN epoch and the transition affects the EOS for total species. We discuss the effect in the EOS on BBN results in Sec.VI.
}.
Then, we can have this relation
\bea 
\dot\rho + 4 H \rho = 0  \label{conservation eq in rde} ,
\eea
after substituting the EOS relation into the conservation equation in Eq.\,(\ref{conservation eq}).
Putting Eqs.\,(\ref{Friedmann eq}) and (\ref{Dot-Friedmann eq}) into 
Eq.\,(\ref{conservation eq in rde}), we obtain\,\footnote{We thank an anonymous referee whose idea is that a direct way of deriving this equation is merely inserting Eqs.\,(\ref{Friedmann eq}) and (\ref{accel eq}) into the EOS, $p- \rho/3 =0 $.}  one equation for $H(t)$
\bea 
 \dot H + 2 H^2 
  + 2 \,\mathcal{C} 
  \bigg( {{d^3  H} \over {dt^3}}  + 7 \ddot{H} H + 4 \dot{ H}^2 +12 \dot H H^2 \bigg)  = 0 .          \label{3rd order}
  \eea
%
%
Note that this is a \textit{third-order} \textbf{\textit{nonlinear}} differential 
equation for $H(t)$.

The Eq.\,(\ref{3rd order}) can be transformed into a unique form with no explicit time dependence \cite{Yun:2022xce}, and its derivative order is reduced by one \cite{Yun:2022xce}:
\bea
&& y + 2 x^2 + 2 \,\mathcal{C} 
\big[  y'' y +   (y')^2  + 7  y' x + 4 y + 12  x^2  \big] y = 0 ~,   
\label{math 2nd order}
\eea 
where $x \equiv H$ and $y \equiv \dot{H}$, and the following relations for $x = x(t)$ 
are utilized 
\bea
&& \ddot x = \dot x { {d  \dot{x} }\over {dx} } ,
 \quad  {{d^3 x} \over {dt^3}} = { {dx} \over {dt}  } { d \ddot{x} \over {dx} }  
                                                =  \dot{x} \Big[ \dot x { {d^2 \dot x} \over { dx^2 } }  +  \Big( { {d  \dot{x} }\over {dx} } \Big)^2      \Big]  ,
\\ &&
 y \equiv \dot x , \quad  y' \equiv { {dy} \over {dx} } =   { {d  \dot{x} }\over {dx} }    = {{ \dot y} \over {y}} ,  
\quad    y'' \equiv { {d^2 y}  \over {dx^2}  }  = {1 \over y^2 } \Big( \ddot{y} - { {\dot y}^2 \over {y}} \Big)   \label{yprime} .
\eea
The \textit{second-order} \textbf{\textit{nonlinear}} differential equation in Eq.\,(\ref{math 2nd order})
will be mainly discussed and used in the rest of this paper.

\section{Standard and non-standard cosmic evolution in the RDE}
\begin{figure*}[]
\centering
\includegraphics[width=0.7\textwidth]{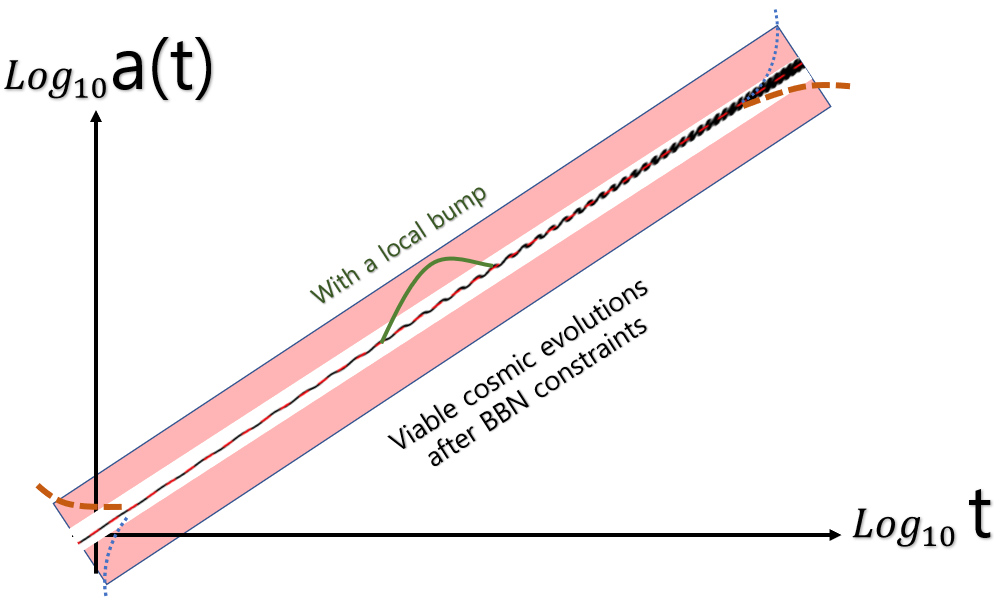}
\caption{Schematically viable non-standard cosmic evolutions constrained the BBN. 
The case of standard cosmic evolution corresponds to a straight line $(a(t) \sim t^{1/2})$ through the oscillating non-standard cosmic evolution (black solid line).
The slightly exaggerated red area indicates 
the small overall size changes compared to standard cosmic evolution (class 1).
The blue dotted and red dashed lines 
stand for allowed deviations from the standard evolution at very early (late) 
time 
of BBN epoch (class 2).
Interestingly (though drawn exaggeratedly), the scale factors with locally 
convex (green solid line) or concave shapes are also viable 
in non-standard cosmic evolutions (class 3).
\label{fig:skematic_scale_factors_on_BBN}}
\end{figure*}

This section discusses both standard and non-standard cosmic evolutions, which 
play an important role in determining the abundances of the light elements.
Unlike Einstein's gravity leading to the standard cosmic evolution $(a(t) \sim t^{1/2})$, 
the MG models can have various forms of non-standard cosmic evolutions in the RDE.
These non-standard cosmic evolutions can be classified into three classes:
1) 
the first class including small changes 
in the standard cosmic evolution;
2) 
the second class containing drastic changes 
over a specific period of standard cosmic evolution;
3) 
the third class involving small local changes in standard cosmic evolution.
All three classes are outlined in Fig.\,\ref{fig:skematic_scale_factors_on_BBN}.

In the first class, two interesting scenarios can be considered.
The first scenario considers the case where non-standard cosmic evolution simply involves 
small overall size changes compared to standard cosmic evolution. 
It is represented by a (slightly exaggerated) red region 
in Fig.\,\ref{fig:skematic_scale_factors_on_BBN}.
The second scenario is the case where the scale factor oscillates.
It is represented by a black line oscillating around the standard cosmic evolution 
of a straight line in Fig.\,\ref{fig:skematic_scale_factors_on_BBN}.
Afterwards, we will discuss the evolution of this oscillating universe in more detail, including BBN.

The second class corresponds to the case where the cosmic evolutions greatly deviate 
from the standard one around a special point in time. 
These are indicated by the blue dotted lines and red dashed lines 
in Fig.\,\ref{fig:skematic_scale_factors_on_BBN}.
Note that the dotted line (dashed line) corresponds to an exponentially increasing or decreasing scale factor (a saturated scale factor).
However, it should be noted that such deviations from the standard cosmology model fail to explain the observations of CMB anisotropies or large-scale structures as well as BBN. 
Therefore, we do not display the evolution results of the non-standard cosmic evolutions in the intermediate period between the early and late BBN as they cannot explain the BBN observation data and the observations of the late universe.

The third class is a case containing a small local deformation of the scale factor.
As an example, this case is shown by the green concave line 
in Fig.\,\ref{fig:skematic_scale_factors_on_BBN}, although this type is not investigated in this study.
In some sense, it refers to new physics associated with the changes in cosmic evolution 
that appear locally at certain points in time, affecting BBN. 
Such a local bump was also shown in the study of BBN with increased high energy tail of the photon \cite{Jang:2018moh}, although the bump stems from not the MG model but the increase of energy density in the standard cosmological model to explain the \ce{^{7}Li} problem. 
We note that the instantaneous increase of the cosmic expansion rate could be advantageous to solve the primordial lithium problem in the models involving a high energetic photon, because it prevents the over-destruction of D element by advancing the freeze-out time of relevant reactions \cite{Jang:2018moh}. However, further microscopic investigation is necessary to verify the change 
of photon spectra in future.

Although the MG models contain rich non-standard cosmic evolutions, 
unfortunately we have found that 
most non-standard cosmic evolutions can not easily satisfy BBN observation data.
In particular, we realize that the primordial abundances of $^4$He and D, 
which have recently been measured most precisely, can strongly constrain these MG models 
and exclude many MG models.
For example, if $\mathcal{C} \gg 1\,{\rm s^2}$ in the eSM, 
the non-standard cosmic evolutions caused 
a change in D abundance of the order of the abundance itself, 
whereas in the case of \ce{^{4}He}, 
this sensitivity was much weaker
; a difference of approximately $O(1 \sim 10)$ occurred.

Here we note that many MG models\,\footnote{
For example, MG models involving $\mathcal{B} R_{ab} R^{ab}$, 
$\mathcal{D}R\Box R$~and/or~$f(R)$,  
where $f(R)$ is a polynomial function of the Ricci scalar $R$.
}
(including the one presented here) in the RDE
generally involve the solution of standard cosmic evolution\,\cite{Yun:2022xce}.
Therefore, it may not be easy to distinguish between the standard cosmological evolution 
model and the MG models using the current BBN observation data.

In the following subsections, we mainly focus on the eSM and 
discuss in detail the standard and non-standard cosmic evolutions in the eSM.
For this purpose, we firstly 
introduce the initial conditions necessary for numerical analysis 
of the \textit{second-order} \textbf{\textit{nonlinear}} differential equation in Eq.\,(\ref{math 2nd order}), 
and then discuss the obtained results in turn.

\subsection{Initial conditions for the second-order \textbf{nonlinear} differential equation describing the evolution of the universe in the RDE}
To obtain a solution for Eq.\,(\ref{math 2nd order}), we need appropriate initial conditions. Since the differential equation is of second order, two initial conditions are required.
For the first initial condition, considering the case of SBBN, 
the earliest time 
of BBN epoch, the initial temperature 
and the standard cosmic expansion rate at that time are used.
Numerically $t_0 \approx 0.15$ ${\rm s}$, $T_0 \approx 2.58 \times 10^{10}$ K, 
$x_0\ (\equiv H) \approx 3.24$ ${\rm s^{-1}}$, and
$ y(x_0)\ \left(\equiv \dot{H}(H)\right) \approx -21$ ${\rm s^{-2}}$ are chosen.
To determine the second initial condition,
one free parameter, $\Delta y'$, which represents
how significantly it  
deviates from the standard value of SBBN, is introduced as follows
\begin{equation}
y' (x) = -4 x ( 1 + \Delta y')~.
\end{equation}
Note that $y'$ is the derivative with respect to $x$, 
not the derivative with respect to time $t$.
Indeed, it is easy to see that $y=-2x^2$ is the simplest solution of Eq.\,(\ref{math 2nd order}) 
that represents standard cosmic evolution.
Numerically, the range of this $\Delta y'$ is  
chosen as follows 
by the sign of $\Delta y'$
\begin{equation}
\Delta y' \in (10^{-15},10^{-3}) ~~~\textrm{for}~~ \Delta y' > 0~~~~~~\textrm{and}~~~~~~
\Delta y' \in (-1,-10^{-15}) ~~~\textrm{for}~~\Delta y' < 0~.
\end{equation}

\subsection{Standard Cosmic Evolution Solution: $a(t) \propto t^{1/2}$ }
\begin{figure}[]
\centering
\includegraphics[width=0.7\textwidth]{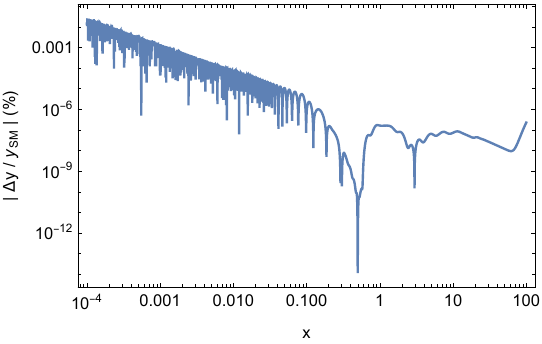}
\caption{In the case of standard cosmological evolution, 
the difference $(\Delta y \equiv y_{\, \rm NS} - y_{\,\rm SM})$ between 
the numerical $(y_{\, \rm NS})$ and analytic ($y_{\,\rm SM}$) solutions 
(based on the analytic solution)
is shown at the percent level.
This reflects that the numerical solution has sufficient precision.
\label{figSM:yVSx_Hvst}
}
\end{figure}
In standard cosmic evolution, the scale factor $a(t)$ is proportional to $t^{1/2}$.
Consequently, cosmic expansion rate $H(\equiv x)$ is $1/(2t)$, 
$y(x)=-2x^2$, and $y^\prime (x)=-4x$. 
For the initial conditions to solve Eq.\,(\ref{math 2nd order}) numerically, 
we take $x=1$, $y=-2$ and $y^\prime = -4$ at $t=1$ second.
It is found that this numerical solution agrees very well with the analytic solution 
in the BBN-related domain of our interest.
To check the accuracy of this numerical solution, we plot 
the difference $(\Delta y \equiv y_{\, \rm NS} - y_{\,\rm SM})$ between 
the analytic solution $(y_{\, \rm SM})$ and 
the numerical solution $(y_{\, \rm NS})$ based on the analytic solution 
in Fig.\,\ref{figSM:yVSx_Hvst}. 
We can see that the numerical solution has sufficient precision 
in the entire range of $x$.

\subsection{Non-Standard Cosmic Evolution Solutions}
In addition to the standard cosmic evolution, various non-standard cosmic evolutions are possible in the eSM.
We thus focus on the non-standard cosmic evolution solutions of Eq.\,(\ref{math 2nd order}), with relevance to the BBN observations discussed later.
\begin{figure}[]
\centering
\includegraphics[width=0.49\textwidth]{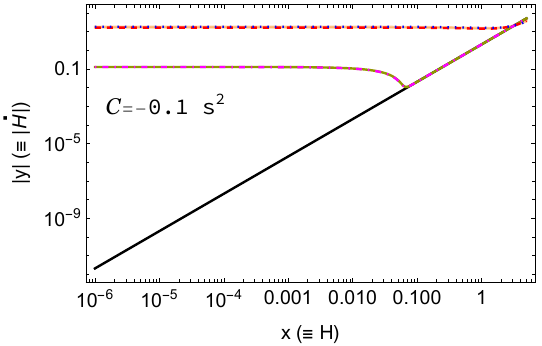}
\includegraphics[width=0.49\textwidth]{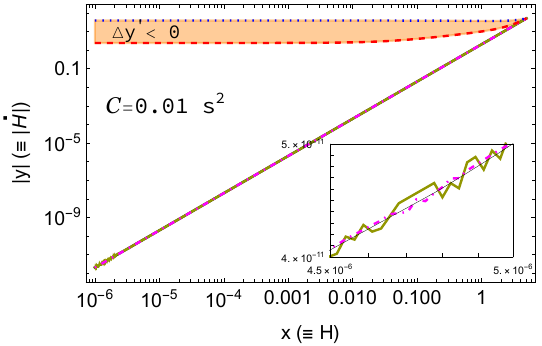}
\includegraphics[width=0.49\textwidth]{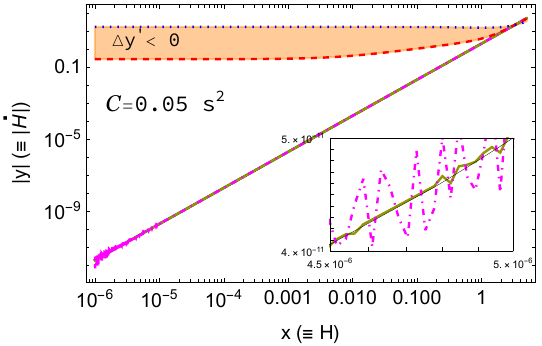}
\includegraphics[width=0.49\textwidth]{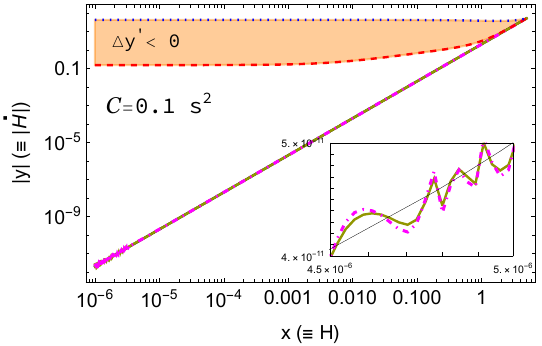}
\caption{Stable solutions of the differential equation in $x\ (\equiv H)$ versus $|y|\ (\equiv |\dot{H}|)$ phase space.
The black solid line in each panel corresponds to the standard cosmic evolution.
Green solid, magenta dot-dashed, red dashed, and blue dotted lines denote stable solutions for $\Delta y'|^{\,+}_{min}$, $\Delta y'|^{\,+}_{max}$, $\Delta y'|^{\,-}_{min}$, and $\Delta y'|^{\,-}_{max}$, respectively.
The orange regions indicate the regions of stable solutions for $\Delta y^\prime < 0$.
We note that there exist two lines (maximum and minimum) 
for each sign of the $\Delta y^\prime$ 
(thus five lines in total including the standard evolution).
Interestingly, when $\mathcal{C}$ is negative, we find no solutions that deviate slightly from the standard evolution, 
whereas when $\mathcal{C}$ is positive, the solutions that oscillate along the standard evolution are found.
From the upper right $(\mathcal{C} = 0.01\, {\rm s^2})$ to the lower right $(\mathcal{C}= 0.1\, {\rm s^2})$,  
the area of allowed solutions increases as the $\mathcal{C}$ value increases.}
\label{figNSEC:Numerical_Solutions_phase_space}
\end{figure}

Eq.\,(\ref{math 2nd order}) is numerically solved with the given two initial conditions, using 
the $\mathtt{NDSolve}$ command in $\mathtt{Mathematica}$.
In particular, in order to find numerically \textbf{\textit{stable}} solutions and overcome the stiffness problem,  
the following options are often used:
\texttt{MaxSteps -> Infinity; Method -> {"StiffnessSwitching", "NonstiffTest" -> False}.}
Depending on the $\Delta y'$ and $\mathcal{C}$ values,
the stable or unstable solutions are found within a given domain.
Among the solutions that retain only numerical stability, 
we investigate four representative solutions, 
$\Delta y'|^{\,+}_{max}, \Delta y'|^{\,+}_{min}, 
\Delta y'|^{\,-}_{max}$, and $\Delta y'|^{\,-}_{min}$, 
which respectively represent the maximum and minimum of the $\pm \Delta y'$.

\begin{figure}
\centering
\includegraphics[width=0.49\textwidth]{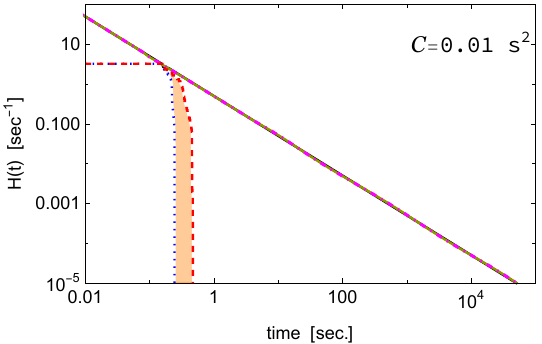}
\includegraphics[width=0.49\textwidth]{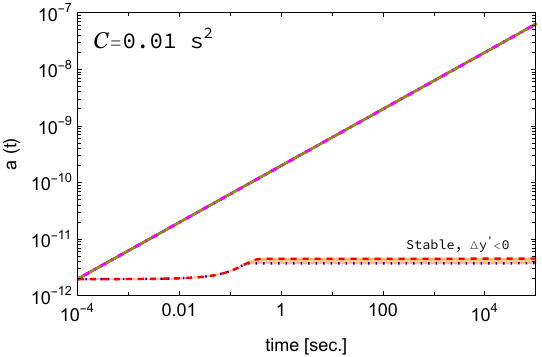}
\includegraphics[width=0.49\textwidth]{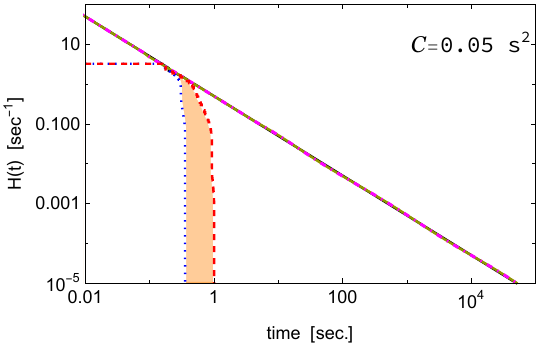}
\includegraphics[width=0.49\textwidth]{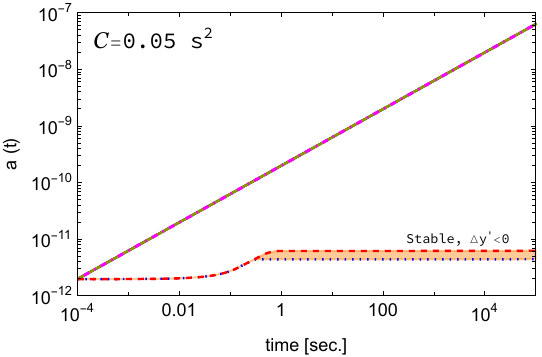}
\includegraphics[width=0.49\textwidth]{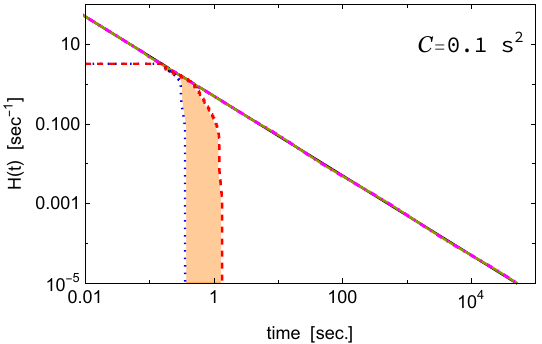}
\includegraphics[width=0.49\textwidth]{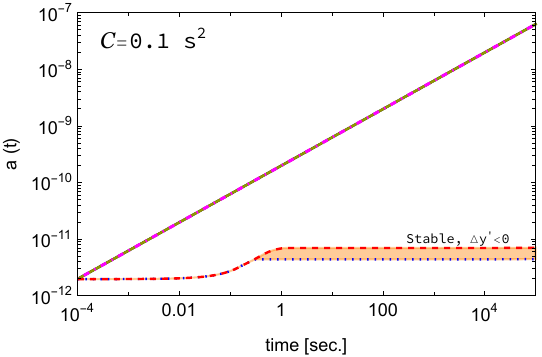}
\caption{
The left (right) panels show the curves of the cosmic 
expansion rate (scale factor) of stable solutions 
for $\mathcal{C}$ = 0.01 ${\rm s^2}$ (top), 0.05 ${\rm s^2}$ (middle), and 0.1 ${\rm s^2}$ (bottom), respectively, as a function of cosmic time. 
Four color lines indicate the cosmic expansion rate with the same $\Delta y'$ denoted in Fig.\,\ref{figNSEC:Numerical_Solutions_phase_space}.
The orange region is the stable solution region 
for $\Delta y^\prime < 0$ 
as shown in Fig.\,\ref{figNSEC:Numerical_Solutions_phase_space}.
We note that, as the $\mathcal{C}$ value increases, 
this orange area gradually increases and 
the cosmic expansion rate drastically decreases after a certain point in time, 
and 
the scale factor almost becomes constant after a period of time. 
}
\label{figNSEC:Numerical_Solutions}
\end{figure}

In Figs.\,\ref{figNSEC:Numerical_Solutions_phase_space} and 
\ref{figNSEC:Numerical_Solutions}, we present two maximum solutions for $\Delta y'|^{\,+}_{max}$ and $\Delta y'|^{\,-}_{max}$ and two minimum solutions for $\Delta y'|^{\,+}_{min}$ and $\Delta y'|^{\,-}_{min}$ with the standard cosmological solution in phase space of $x\ (\equiv H)$ and $|y| \ (\equiv |\dot{H}|)$.
In addition, the following $\mathcal{C}$ values are considered in order to see the changes in the  
numerical solutions: $\mathcal{C} \in \{-0.01, 0.01, 0.05, 0.1, 10, 10^4, 10^7\}$ ${\rm s^2}$.
It turns out that when $\mathcal{C}$ is greater than 0.1 ${\rm s^2}$, 
the amount of primordial deuterium varies enormously due to the significant change in the cosmic expansion rate, 
which excludes the solutions with $\mathcal{C} > 0.1\, {\rm s^2}$.
(see Sec.\,VI 
for more detailed discussion).
For this reason, 
only the cases of $\mathcal{C}$=-0.1 (top left), 0.1 (top right), 0.05 (bottom left), 
and 0.1 (bottom right) $\rm s^2$ are presented in Fig.\,\ref{figNSEC:Numerical_Solutions_phase_space}.
The orange areas with the blue dotted line ($\Delta y'|^{\,-}_{max}$) 
and red dashed line ($\Delta y'|^{\,-}_{min}$) as boundaries 
represent the allowed region of 
stable solutions
for $\Delta y' < 0$. 
These areas are obtained through visual inspection according to the change of $\mathcal{C}$ value and the change of boundary conditions.
When $\mathcal{C} > 1\,{\rm s^2}$, the area corresponding to $\Delta y' > 0$ begins to appear 
noticeably (see Appendix A.).

For the top left panel 
corresponding to $\mathcal{C}=-0.1$ $\rm s^2$, we find that both stable solutions for positive and negative $\Delta y'$ significantly deviate from the standard cosmological evolution. Such deviations also lead to a significant change in the freeze-out time of thermonuclear reactions in BBN, resulting in inconsistencies between the predicted primordial abundances and observations.
Therefore, there are no numerical solutions that simultaneously satisfy the numerical stability 
and BBN observation data.
Moreover, even if the $\mathcal{C}$ value is 
within $-0.1\, {\rm s^2} < \mathcal{C} < 0$, 
no solution exists that satisfies the BBN observation data.
Considering these results and the changes in the solutions according to 
their $\mathcal{C}$ dependence,
we found that the free parameter $\mathcal{C}$ of the eSM can not be negative if we demand consistency with the BBN observation data.

As the value of $\mathcal{C}$ changes from 0.01 ${\rm s^2}$ to 0.1 ${\rm s^2}$, 
we can see in Fig.\,\ref{figNSEC:Numerical_Solutions_phase_space} that 
the stable solution region (orange region) of $\Delta y'<0$ gradually increases.
However, once again, it should be noted that these kinds of solutions represent 
a stationary universe in which the scale factor does not change 
after a certain point in the RDE, which fails to explain observed primordial abundances. 
(see the orange areas in the right panels of Fig.\,\ref{figNSEC:Numerical_Solutions}).
Consequently, only numerical solutions with $\Delta y' > 0$ oscillating along the 
diagonal (i.e., standard cosmic evolution, black solid line) in the $\mathcal{C} > 0$ panels 
are likely to fit the BBN observations.
The green solid line ($\Delta y'|^{\,+}_{max}$) 
and magenta dot-dashed line ($\Delta y'|^{\,+}_{min}$) 
representing the oscillating numerical solutions 
on the black solid lines are shown in Fig.\,\ref{figNSEC:Numerical_Solutions_phase_space}.
In addition, the small figures are inserted into the positive $\mathcal{C}$ panels 
to show the actual oscillations of the solutions 
that are not easily visible on the log scale.
We also observe that these oscillating solutions always exist
when $\mathcal{C}$ is positive and less than $10^7$ ${\rm s^2}$.

In Fig.\,\ref{figNSEC:Numerical_Solutions} we present the 
cosmic expansion rates 
(left) and the scale factors 
(right) 
for the three cases $\mathcal{C}$=0.01 (top), 0.05 (middle), and 0.1 (bottom) ${\rm s^2}$.
As in the previous case, overlapping diagonal lines correspond to the oscillating 
non-standard cosmic evolutions along standard cosmic evolution.
The orange areas 
correspond to the orange areas in the previous 
Fig.\,\ref{figNSEC:Numerical_Solutions_phase_space}.
As before, it can be seen that as the $\mathcal{C}$ value increases, 
the allowed orange area also increases.
Interestingly, these non-standard cosmic evolutions can be thought of 
as describing a stationary universe in which the cosmic expansion rate decreases 
very rapidly after a certain point in time and the corresponding scale factor 
remains constant.
However, for these non-standard cosmic evolutions, 
it is 
hard to satisfy the known BBN observational data, 
as can be easily expected.

\begin{figure}[]
\centering
\includegraphics[width=0.49\textwidth]{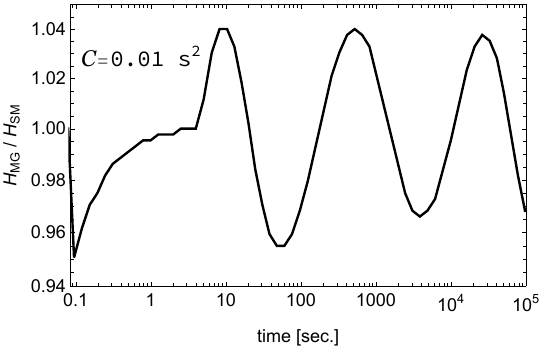}
\includegraphics[width=0.49\textwidth]{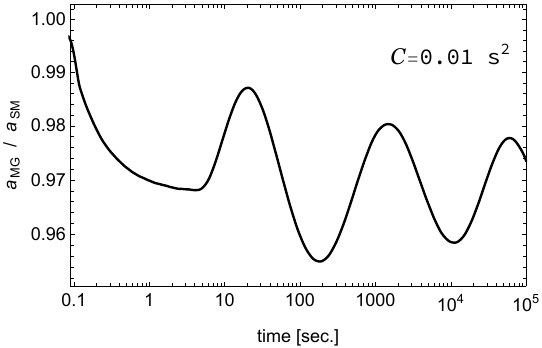}
\caption{
The left (right) panel shows the ratio between the two cosmic expansion rates 
(the scale factors)  
by standard and non-standard cosmic evolutions. For non-standard cosmic evolution, 
the oscillating case with $\mathcal{C} = 0.01$ ${\rm s^2}$ and {$\Delta y'=10^{-13}$} is selected. 
The cosmic expansion rate (scale factor) in the eSM is expressed as $H_{MG}(a_{MG})$.  
For the standard case, it is expressed as $H_{SM}$($a_{SM}$).
Both panels confirm that this non-standard cosmic evolution is oscillating 
with a difference of about a few percent compared to the standard cosmic evolution.
Note the slight asymmetry in the maximum amplitude in 
the ratio of the cosmic expansion rates.
}
\label{fig:Ratios}
\end{figure}
The left (right) panel of Fig.\,\ref{fig:Ratios} shows the ratio between 
the cosmic expansion rates (the scale factors) in standard and non-standard cosmic evolutions. 
The case of \textbf{\textit{oscillating cosmic evolution}} 
corresponding to $\mathcal{C}=0.01$ ${\rm s^2}$ is used to obtain these ratios.
The $H_{MG}\ (H_{SM})$ and $a_{MG}\ (a_{SM})$ denote the cosmic expansion rate and scale factor 
in non-standard (standard) cosmic evolution, respectively.
For the cosmic expansion rates, we can see that 
there is an asymmetrical difference of about $-5\% \sim +4\%$ 
along the oscillations of the ratio.
In the case of the scale factor, 
the net negative contribution is generated due to the asymmetric oscillation ratio of the cosmic expansion rate.
It is confirmed that this negative contribution can reach up to $-5.5\%$, and 
almost similar results can be obtained when $\mathcal{C}=0.05$ ${\rm s^2}$ and $\mathcal{C}=0.1$ ${\rm s^2}$.

\section{Input values used for the \texttt{PRIMAT} BBN calculations}
The \texttt{Mathematica} code, \texttt{PRIMAT}, is used to calculate 
the primordial abundances in the \textbf{\textit{oscillating cosmic evolution}} of the eSM.
The evolution of the universe 
by the usual Friedmann equations has been modified 
within this code to account for this MG effect. Specifically, we obtained the cosmic expansion rate by numerically solving Eq.\,(\ref{math 2nd order}) and embedded the interpolated solutions into the PRIMAT code.
The important constants adopted in the calculations are introduced in order as follows.
For the masses of Z boson and W boson, $m_{\rm Z} = 91.1876$ GeV and $m_{\rm W} = 80.385$ GeV 
are used, respectively. 
The values of $\cos \theta_C = V_{ud} = 0.97420(20)$ and $g_A = C_A /C_V = 1.2723(23)$ 
are taken from the Particle Data Group (PDG) \cite{ParticleDataGroup:2016lqr}.
The Fermi constant $G_F = 1.1663787 \times 10^{-5}$ $\rm{GeV}^{-2}$ and 
the fine structure constant $\alpha_{\rm FS} = 1 / 137.03599911$ are used. 
The neutron lifetime ($\tau_n$) given by Ref.\,\cite{Serebrov:2017bzo},
$879.5 \pm 0.8\,{\rm s}$ is used\footnote{
This value is slightly larger than 
the weighted average $\tau_n = 878.4 \pm 0.5\,{\rm s}$ of the PDG\,\cite{Workman:2022ynf}.
However if we include the result from the in-beam measurement with a comparable error,
the PDG value becomes $\tau_n = 879.6 \pm 0.8\,{\rm s}$\,\cite{Workman:2022ynf}.
This value is very similar to the value used in 
\texttt{PRIMAT}\,\cite{Pitrou:2018cgg,Pitrou:2020etk}.
}. 
The baryon-to-photon ratio today $(\eta_0)$ is $6.09 \times 10^{-10}$, 
which is corresponding to the baryon density by the 
Planck observation of the cosmic microwave background, 
$\Omega_b\, h^2 = 0.022250 \pm 0.00016$\,\cite{Planck:2015fie}
, where $h=H_0/H_{100}=0.6727 \pm 0.0066$.

\section{Observed Light element abundances} \label{sec.V}

The primordial abundances are extracted from observations of cosmological objects 
with low metallicity.
Using their galactic chemical evolution, pristine abundances 
have been accumulated from different objects. 
This section briefly introduces these observations in order.

The primordial \ce{^{4}He} content is estimated from observations of the metal-poor 
extragalactic H\texttt{II} (ionized hydrogen) regions.
For each H\texttt{II} region, the \ce{^{4}He} abundance is determined in the model 
to reproduce the observed emission lines, and then extrapolated to zero metallicity to account 
for stellar formation.
Here we use the recent determination of \ce{^{4}He} mass fraction as ${\rm Y_p} = 0.2449 \pm 0.0040$~\cite{Aver:2015iza}.
When the central values of neutron lifetime, the baryon-to-photon ratio, and 
the reaction rate adopted are taken, 
the results $({\rm Y_p} = 0.2472)$ of the SBBN model agree well within 
the observational uncertainty of 1$\sigma$.

The abundance of primordial D is estimated with observations of 
metal-poor Lyman-$\alpha$ absorption systems in the foreground of quasistellar objects.
Until recently, a significant scatter were shown in the distribution of D/H 
observations~\cite{Pettini:2012}, but recently new observations and reanalysis of 
previous data~\cite{Cooke:2013cba,Cooke:2016rky,Cooke:2017cwo,Balashev:2015hoe,Riemer-Sorensen:2017pey} 
have resulted in a plateau with a very small scatter.
Hence, we use the recommended value provided by Ref.\,\cite{Cooke:2017cwo}:
D/H $=(2.527 \pm 0.030) \times 10^{-5}$. 
Note that the central value is lower and has less uncertainty 
compared to the previous determinations.
Also, since the value of $2.438 \times 10^{-5}$ predicted by SBBN 
does not fit the observed one within 2$\sigma$, 
the 4$\sigma$ region is considered in the following discussion.

The \ce{^{3}He} abundances are measured via the 8.665 GHz hyperfine transition of 
the \ce{^{3}He^+} ion in the H\texttt{II} regions of the galaxy.
These are not primordial quantities, but present values deduced from galactic chemical evolution 
that takes into account the destruction and creation of nucleons in the stars.
Thus, unlike \ce{^{4}He}, the evolution of this quantity over time is not precisely 
known~\cite{Vangioni-Flam:2002cvh}.
We take this \ce{^{3}He}/H $ < (1.1 \pm 0.2) \times 10^{-5}$ as the upper bound~\cite{Bania2002}.

In the case of primordial \ce{^{7}Li}/H, the abundance is estimated from observations of the 
galactic metal-poor stars. After BBN, it is produced inside the star at temperatures 
as low as 2.5 MK, but can also be easily destroyed by proton capture.
Here we adopt the following analysis of Sbordone et al. \cite{Sbordone:2010}, namely
$\ce{^{7}Li}/{\rm H} = (1.58 \pm 0.3) \times 10^{-10}$.

\section{Results}
\subsection{BBN results from oscillating cosmic evolutions}
\begin{table}
\caption{
BBN abundances in observational data and model calculations, and differences between the two values.
The first column denotes the models, where SBBN (eSM) means the SBBN model (extended Starobinsky Model). 
The 2nd column is the ${\rm Y_p}$ and 
the 3rd column is the relative difference from the observation.  
The 4th column is the relative abundance of D and 
the 5th column for the relative difference.
Similar to the fourth and fifth columns, 
the sixth (eighth) column shows \ce{^{3}He/H} (\ce{^{7}Li}/H), and 
the seventh (ninth) column shows their relative difference, respectively.
In the case of observation and the SBBN model, 
an uncertainty of $1\sigma$ was taken into account.
In the eSM with $\mathcal{C}=0.01$ ${\rm s^2}$, the theoretical uncertainty of each element 
is calculated directly to compensate for the effects of our model's simple 
assumptions. 
(see the footnotes on page 5 and the text on page 14 for details). 
}
\label{tab:1}
\begin{adjustbox}{width=\textwidth}
\begin{tabular}{c | l c | l c | l c | l c}
\hline 
\hline
~~ ~~ & ~~${\rm Y_p}$~~ & ~~$\Delta {\rm Y_p}$~~ & ~~$10^5\times$ D/H~ & ~$\Delta($D/H$)$~ & 
~~$10^5\times $\ce{^{3}He}/H~~ & ~$\Delta$(\ce{^{3}He}/H)~ & 
~~$10^{10}\times $\ce{^{7}Li}/H~~ & ~~$\Delta($\ce{^{7}Li}/H)~ \\ 
Observations~~ &~ $0.2449 \pm 0.0040$~~ & 0 &~ $2.527 \pm 0.030$~~ & 0 
&~ $< 1.1 \pm 0.2$~~ & 0 &~ $1.58 \pm 0.3$~~ & 0 \\ 
\hline
\hline\
Model SBBN$(\mathcal{C}=0\, {\rm s^2})$ &~ $0.24709 \pm 0.00017$ & $+0.0023$ ~&~ $2.459 \pm 0.036$ & $-0.089$ 
&~ $1.074 \pm 0.026$ 
& $-0.07$ &~ $5.623 \pm 0.247$ & $+3.92$ \\ 
\hline
\hline 
~~eSM$(\mathcal{C}=0.01\, {\rm s^2})$ &~  $0.2287 \,\pm\, 0.0229$ & 
$-0.0162$ ~&~ $1.912 \,\pm\, 0.115$ & 
$-0.615$ &~ $0.96 \,\pm\, 0.02$ & 
$-0.14$ &~ $6.48 \,\pm\, 0.32$ & 
$+4.90$ \\ 
\hline 
\hline
~~eSM$(\mathcal{C}=0.05\, {\rm s^2})$ ~&~ $0.2287$ & 
$-0.0162$ &~ $1.913$ & $-0.614$ &~ $0.96$ & $-0.14$ &~ $6.48$ & $+4.90$ \\ 
\hline 
~eSM$(\mathcal{C}=0.1\, {\rm s^2})$ ~&~ $0.2287$ & 
$-0.0162$ &~ $1.914$ & $-0.613$ &~ $0.96$ & $-0.14$ &~ $6.48$ & $+4.90$ \\ 
\hline 
\hline
\end{tabular} 
 \end{adjustbox}
\end{table}
\begin{figure}[]
\centering
\includegraphics[width=1\textwidth]{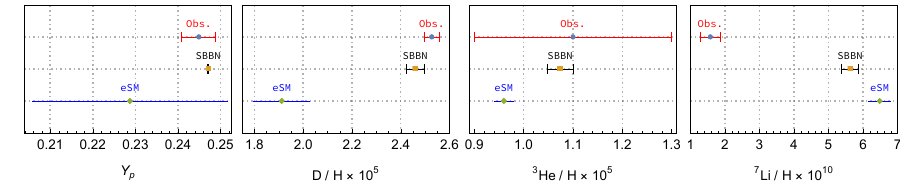}
\caption{Primordial ${\rm Y_p}$, D/H, \ce{^{3}He}/H,  \ce{^{7}Li}/H 
in observations (red) and predictions by the SBBN (black) and eSM (blue). 
The error bars represent the central values and their uncertainties in each model.
The error bar of eSM is expressed without fences 
to emphasize that this error bar is the theoretical uncertainty.
\label{fig:AllYields}}
\end{figure}  
Table \ref{tab:1} shows the BBN abundances in observations and models,
where the $\rm Y_p$, D/H, \ce{^{3}He}/H, and \ce{^{7}Li}/H 
are presented.  
Differences $(\Delta)$ between the observed values and the predicted values of 
the given model are also shown in the 3rd column $(\Delta {\rm Y_p})$, 
fifth $(\Delta  $D/H), seventh $(\Delta $\ce{^{3}He}/H), and ninth $(\Delta $\ce{^{7}Li}/H), respectively.
The observed  central values and 
their $1\sigma$ uncertainties, mentioned in Section V, are presented in the second line of the first row.
By comparing the observations with the predictions of the SBBN model, 
the $\rm Y_p$  and \ce{^{3}He}/H fall within $1 \sigma$ and in the case of D/H, 
within $3 \sigma$,
but in the case of \ce{^{7}Li}/H, they are off by more than $10\sigma$ from each observational uncertainty.
In particular, the central value of the \ce{^{7}Li}/H is approximately 3.6 times larger 
than the observational one.

For the eSM, three benchmark models with $\mathcal{C} = 0.01\,{\rm s^2}, 0.05\,{\rm s^2}$, and $0.1$ ${\rm s^2}$ 
are selected\,\footnote{
In the models with $\mathcal{C} \ge 1 $ ${\rm s^2}$, we find that
the expected amount of deuterium is very different from the observed value.
}
\footnote{
We find that in these three benchmark models, the results of the model do not significantly depend on changes in $\Delta y'$. This is because the oscillation amplitude is not sensitive to the changes in $\Delta y'$. Hence, for the model with $\mathcal{C} = 0.01\,{\rm s^2}$, we have chosen $\Delta y' = 10^{-13}$.} 
.
We observe that the previous simple assumptions of EOS and energy density 
discussed below Eq.\,(\ref{Dot-Friedmann eq}), 
which are used to derive 
the \textit{second-order} \textbf{\textit{nonlinear}} differential equation, 
actually affect the BBN abundances, and the resulting impact can not be neglected.
Using the SBBN model, it is checked that 
in the case of \ce{^{3}He}/H, a difference of about 2$\%$ is produced, and 
for the \ce{^{7}Li}/H and D/H, differences of $5\%$ and $6\%$ occur, respectively. 
Unfortunately, for the $\rm Y_p$, a large difference of about $10\%$ occurs 
because of the precise data.
To reflect the impact on the eSM, 
we treat the differences as theoretical uncertainties for the light elements. 
The calculated theoretical uncertainty in the eSM with $\mathcal{C}=0.01$ ${\rm s^2}$  is shown
along with the central value of each element in the fourth row of Table \ref{tab:1} 
and in Fig.\,\ref{fig:AllYields}.
Note that these theoretical uncertainties are different from those $(1\sigma)$ 
between observations and in the SBBN model.
To emphasize this point, in the case of the SBBN model and observations, 
the error bars with fences at the error limits are presented, 
while in the case of eSM, they are displayed as open error bars in Fig.\,\ref{fig:AllYields}.

First, looking at the results from the 3rd ($\mathcal{C}$=0.01 ${\rm s^2}$ ) to 5th ($\mathcal{C}$=0.1 ${\rm s^2}$ ) rows 
of Table 1, it can be seen that the expected BBN abundances are almost the same 
despite the difference in $\mathcal{C}$ values.
Although the central value of ${\rm Y_p}$ predicted by the eSM is excluded by the observational data at the $4\sigma$ level, whether the predicted value of ${\rm Y_p}$ with the eSM is consistent with the observational data depends on its theoretical uncertainty.
Therefore, it seems reasonable not to use $\rm Y_p$ to validate these benchmark models
unless this theoretical uncertainty is reduced to the level of the observational error.
To further reduce the theoretical uncertainty, 
it is necessary to improve the approximated EOS used in this paper 
for computational simplicity. 
We leave this rigorous treatment as a future paper.
On the other hand, in the case of \ce{^{3}He}/H, it is included within $1 \sigma$ of 
the observed one due to its large error range, and in the case of \ce{^{7}Li}/H, 
it is out of the observed value by about $14 \sigma$ or more.
In the case of D/H, the central value decreases by about $22\%$ 
compared to the SBBN one, which indicates that it is about $20 \sigma$ away from the center of the observed value. 
However, since all models contain very small uncertainties in the case of deuterium,
there may be a potential to distinguish eSM from SBBN 
as the precision of deuterium increases in the near future.

As shown in Sec.\,III, three benchmark models 
oscillate by a few percent difference compared to the standard cosmic evolution.
For example, in the benchmark model with $\mathcal{C}=0.01\,{\rm s^2}$,
the cosmic expansion rate oscillates with a difference of $+4 \sim -5\%$ 
compared to the standard evolution (see the left panel in Fig.\,\ref{fig:Ratios}), 
and the scale factor oscillates with a smaller amplitude overall
(see the right panel in Fig.\,\ref{fig:Ratios}).
By considering these oscillating effects in the eSM, 
the BBN results of these benchmark models can be qualitatively understood 
as follows.

First, at $T \gtrsim 10^{10}\,{\rm K}$, 
the neutron-to-proton ratio in equilibrium satisfies the condition of $n/p \simeq \exp(-\Delta m/T)$ where $\Delta m$ denotes the mass difference between neutron and proton. The slow cosmic expansion rate leads to the later freeze-out of this weak interaction rate, so $n/p$ decreases. Since the neutron is the main seed to produce the $^4$He, the reduced neutron abundance decreases the ${\rm Y_p}$ at the He synthesis stage.

Second, at $T \lesssim 10^{9}\,{\rm K}$, the reduced $n/p$ due to the slow expansion rate decreases the main production process of D, ${^1}{\rm H}(n,\gamma){\rm D}$. In addition, the slow expansion rate delays the freeze-out of reactions of ${\rm D}({\rm D},n){^3\rm He}$ and ${\rm D}({\rm D},p){^3\rm H}$, which are main reactions destroying the D. Therefore, the slower expansion rate reduces the D/H. Since the D is also a source to produce ${^3}$H and ${^3}$He, the reduced D/H decreases the ${^3}$H/H and ${^3}$He/H. We note that the ${^3}$H decays into the ${^3}$He so that the final abundance of ${^3}$He includes the abundance of the ${^3}$H.

Third, in the case of \ce{^{7}Li}/H in the SBBN model, the prediction is 
about 3--4 times higher than the observations. 
Although one proposed possible solutions exploiting long-lived negative charged massive particles \cite{Kusakabe:2014moa} or Tsallis statistics for photon distribution \cite{Jang:2018moh}, these solutions involve additional hypotheses requiring the experimental or observational cross-verification.
Moreover, the electron screening effect, modification of decay lifetime, 
and updated reaction rates \cite{Cyburt:2008up,Cyburt:2008et}
do not seem to be of much help 
to solve this $\ce{^{7}Li}$ problem.
Often the updated reaction rates exacerbate the situation \cite{Cyburt:2008et}, 
and many new solutions or models are ruled out by the improved precision on
D/H observations \cite{Cooke:2017cwo}.

The \ce{^{7}Li} is mainly produced via $^3$H($\alpha,\gamma$)$^{7}$Li and 
destroyed via \ce{^{7}Li}(p,$\alpha$)\ce{^{4}He}.
Therefore, the \ce{^{7}Li} abundance is smaller because of the smaller amount of $^3$H.
The \ce{^{7}Be} is mainly produced via $^{3}$He($\alpha$, $\gamma$)$^{7}$Be and 
destroyed via \ce{^{7}Be}(n,p)\ce{^{7}Li}.
Although the reduced \ce{^{3}He} abundance slightly reduces the formation of \ce{^{7}Be}, 
the greatly reduced neutron abundance significantly decreases the destruction rate of \ce{^{7}Be}.
As a result, the abundance of \ce{^{7}Be} becomes relatively large.
The increased \ce{^{7}Be} is converted to \ce{^{7}Li} through an electron capture process 
after BBN and recombines with electrons.
Since the abundance of \ce{^{7}Be} is about 18 times greater than 
that of \ce{^{7}Li} in SBBN, the \ce{^{7}Be} abundance significantly accounts for the final abundance of \ce{^{7}Li}.
Thus, a slower expansion rate leads to a higher primordial \ce{^{7}Li} abundance.

\begin{figure}[]
\centering
\includegraphics[width=1.0\textwidth]{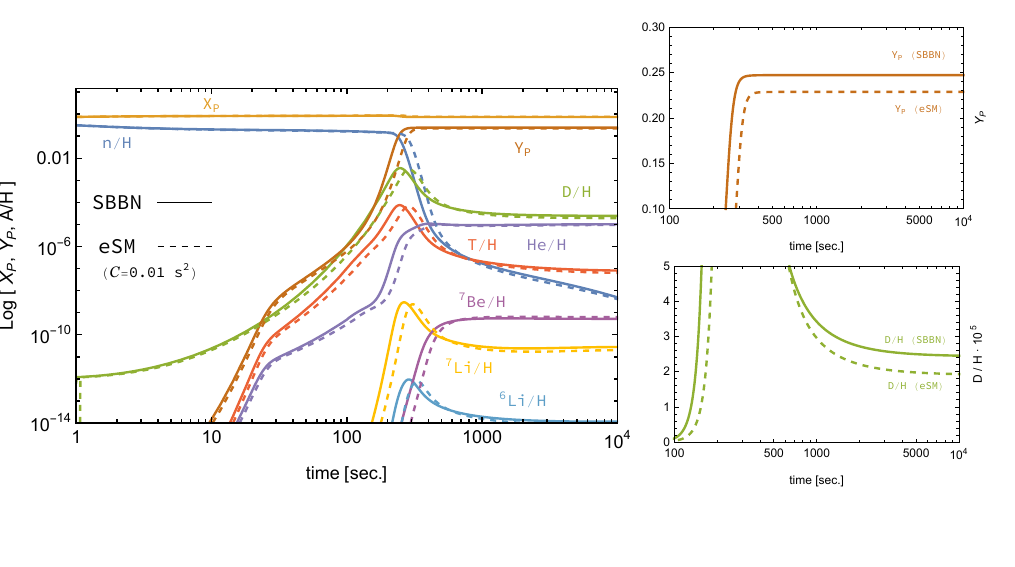}
\vspace{-2cm}
\caption{Evolutions of light elements abundances in the SBBN (solid lines) and 
in the eSM with $\mathcal{C} = 0.01\,{\rm s^2}$ and $\Delta y'=10^{-13}$ (dashed lines).
Noticeable differences in these light elements are shown despite the logarithmic scale.
In addition, the abundant $\rm{Y_{\rm p}}$ (D/H) is presented in a linear scale 
on the upper (lower) right panel to show in more detail. 
For temperature information over time for both models, 
see Appendix B: Time-temperature relation.
\label{fig:LightYields}}
\end{figure}  
Fig.\,\ref{fig:LightYields} shows the evolutions of the light elements abundances 
as a function of time. 
The solid (dashed) lines correspond to the predicted abundances of the SBBN (eSM) model.
For the eSM, the model with $\mathcal{C}=0.01$ ${\rm s^2}$ is chosen.
Additionally, we plot the evolution of the rich $\rm{Y_{\rm p}}$ (D/H) 
on a linear scale in the upper (lower) right panel to show the final number of abundance 
in more detail.
Note that these $\rm{Y_{\rm p}}$ and D/H are the most important elements 
in constraining the free parameter $\mathcal{C}$ because they are abundant 
and the most precisely measured.
From this, once again, 
it may be expected that the precise primordial abundance of $\rm{Y_{\rm p}}$ or D/H 
could be the best criterion for distinguishing between the eSM from the SBBN.


\subsection{Constraints of the $\mathcal{C}$ parameter for the eSM that fits the BBN observation results}

\begin{figure}[]
\centering
\includegraphics[width=0.49\textwidth]{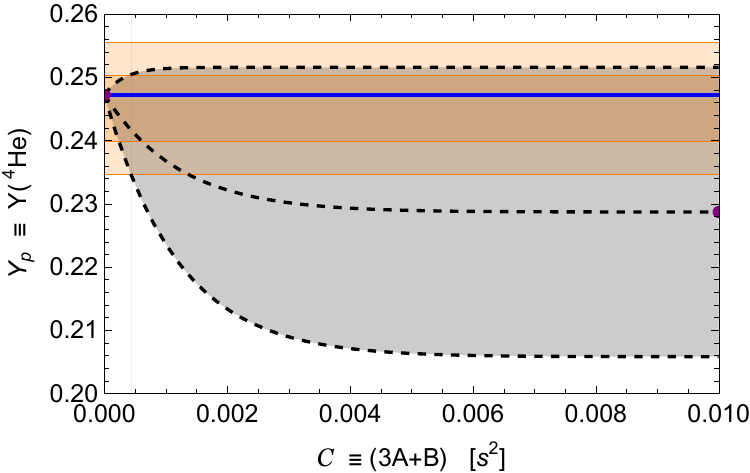}
\includegraphics[width=0.49\textwidth]{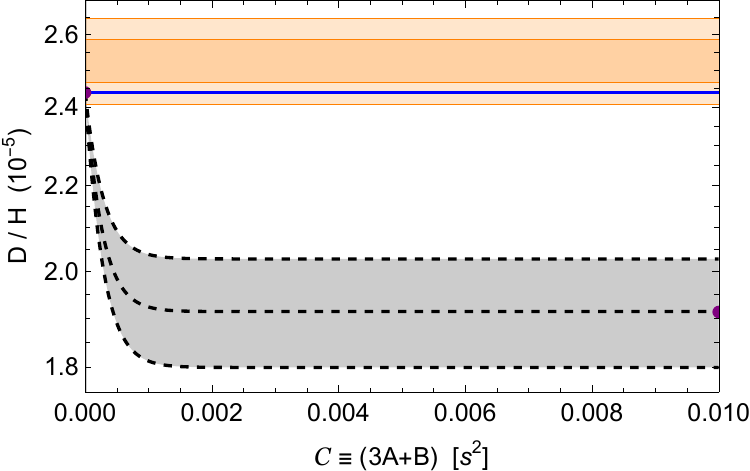}
\caption{The final abundances of D/H (left) and ${\rm Y_p}$ (right) as a function of $\mathcal{C}$.
The light (dark) orange region corresponds to the region corresponding to $4\sigma\ (2\sigma)$ of 
the observed values.
The blue solid line represents the calculated center value in the SBBN model and 
the black dashed lines represent the interpolated values in the eSM. 
Three black dashed lines in each panel indicate the interpolated lines using the upper, central, and lower values of predictions with theoretical uncertainties in Table \ref{tab:1}, respectively.
Consequently, $\mathcal{C}$ is constrained as $\mathcal{C} < 4.3 \times 10^{-4}\,{\rm s^2}\ (1.3 \times 10^{-5}\, {\rm s^2})$ for ${\rm Y_p}$ (D/H) within $4\sigma$ range.
Compared to ${\rm Y_p}$, 
the case of deuterium constrains the free parameter $\mathcal{C}$ of non-standard cosmic evolution models more strongly. 
\label{fig:MG_C_constraints}
}
\end{figure}

In this subsection, we investigate the specific conditions of the parameter $\mathcal{C}$ 
under which BBN observations and predicted values within the eSM 
can be consistent with each other.
For this purpose, interpolation is performed using the abundances 
of $\rm Y_p$ and D/H at $\mathcal{C}=0.01\,{\rm s^2}$ (eSM) and $\mathcal{A}=\mathcal{B}=0$ (SBBN).
Note that compared to the other elements, 
these two elements have precisely measured observational data 
(see the Sec.\,\ref{sec.V}) and 
are sensitive to the value of parameter $\mathcal{C}$.
Fig.\,\ref{fig:MG_C_constraints} shows the changes in the amount of $\rm Y_p$ (left panel) and 
D/H (right panel) according to the $\mathcal{C}$ value in the eSM.
In each panel, the black dashed lines represent the interpolated values in the eSM, 
and the blue solid line represents the central value of the SBBN model.
The dark (light) orange regions correspond to the observation results 
in $2\sigma\ (4\sigma)$ of each element.
In the case of $\rm Y_p$, the SBBN central value agrees well within $2 \sigma$, 
while in the case of D/H, it is located between $2 \sigma$ and $4 \sigma$.
In both panels, we obtain the constrained $\mathcal{C}$ value as $ 4.3 \times 10^{-4}\,{\rm s^2}$ ($1.3 \times 10^{-5}\,{\rm s^2}$) for $Y_p$ (D/H) within $4\sigma$ range.
\footnote{
It is worth noting that this result provides broad upper bound compared to 
the one obtained from previous studies on neutron stars. 
For example, according to Ref.\,\cite{Arapoglu:2010rz}, the model of $f(R) = R + \alpha R^2$ that includes only $R^2$ term is constrained from the mass-radiation relation with observed neutron stars, which leads to $\alpha \sim 10^{9}\,{\rm cm}^2$. On the other hand, our result constrained by the BBN observations corresponds to the order of $\mathcal{C} \sim 10^{-5} c^2 {\rm s^2} \sim 10^{16}\,{\rm cm}^2$. Consequently,
there is an order difference of $10^7$ between the constraint value presented 
in Ref.\,\cite{Arapoglu:2010rz} and our value, if we assume $\mathcal{C} \simeq \alpha$.
}
We can thus conclude that precise observations of these two elements strongly constrain 
the eSM, 
and in other words non-standard 
\textit{\textbf{oscillating cosmic evolution}} cannot deviate too much 
from standard evolution.
Furthermore, it can be confirmed that the D/H abundance gives a stronger constraint 
than that of $\rm Y_p$, which is consistent with the fact that the D/H abundance is most sensitive to the $\mathcal{C}$ parameter of the eSM, as noted above.

As previously mentioned, the small scale factor oscillations obtained in this study are sufficiently small that they would rapidly decay during the RDE \cite{Yun:2022xce}. Consequently, the constraint on $\mathcal{C}$ derived from the BBN observations may not be discernible in the CMB or large-scale structure formation. Nevertheless, more precise BBN observations are expected to yield more accurate information on $\mathcal{C}$. Therefore, it is desirable to investigate cosmic evolution with rigorous calculations that involve the exact equation of state and various initial conditions.

\section{Summary and Conclusion}
In this study, we considered the eSM obtained by adding one $R_{ab}R^{ab}$ term 
to the Starobinsky model.
In particular, beyond the Einstein-Hilbert action, 
the effects of higher-order terms $R^2$ and $R^{ab}R_{ab}$ are examined in the RDE.
While the previous studies were mainly based on the standard cosmic evolution, 
which follows $a(t) \sim t^{1/2}$, 
this study focuses on the non-standard cosmic evolutions 
generated by the effects of the higher order terms.
The \textit{second-order} \textbf{\textit{nonlinear}} differential equation 
describing the interesting cosmic evolutions is introduced 
after some simple assumptions about the EOS and energy density in the RDE.
In this \textbf{\textit{nonlinear}} differential equation, 
the coefficients ($\mathcal{A}$, $\mathcal{B}$) of 
the two higher-order terms ($R^2$, $R^{ab}R_{ab}$) are grouped together and 
represented by a single parameter $\mathcal{C}$ $(\equiv 3\mathcal{A}+\mathcal{B})$.
To solve the equation numerically, 
the $\tt{NDsolve}$ command with suitable options in \texttt{Mathematica} is used.
Different types of numerically stable solutions 
corresponding to standard and non-standard cosmic evolutions have been found.
We have thus demonstrated that they really involve 
the \textit{oscillating, diverging}, or \textit{stationary} 
non-standard cosmic evolutions in the RDE.

We also tracked down the phenomenologically viable eSM considering BBN, 
the most important observation in the RDE.
It has been discussed that 
\textit{divergent} and/or \textit{stationary} non-standard cosmic evolution models 
 can not easily fit these BBN observations.
Therefore, the models with \textit{\textbf{oscillating non-standard cosmic evolution}} 
are mainly studied in this paper and 
three benchmark models with small $\mathcal{C}$ values are selected.
The primordial abundances of light elements are then calculated in these models, and 
their BBN prediction results are compared and discussed 
with those of observation and the SBBN model.
To compensate for the impact of simple assumptions about EOS and energy density, 
we additionally set the theoretical uncertainties of approximately $2 \sim 10\%$ levels
in the expected abundances of the light elements.
For the $\rm Y_p$, the observed $1\sigma$ interval is included within 
the predicted $1\sigma$ interval, whereas for the \ce{^{3}He}, 
it is vice versa.
In the case of D/H and \ce{^{7}Li}, the predicted $1\sigma$ intervals do not overlap the 
observed $1\sigma$ interval and are very far from the observational data.
Comparing only the predicted central value with the observed central value, 
the D/H abundance becomes $24\%$ smaller and 
the \ce{^{7}Li} abundance increases by $410\%$.
By further observing the change in each abundance while changing the $\mathcal{C}$ value, 
we further find that the D/H abundance is the most sensitive to the change of 
the $\mathcal{C}$ value compared to other abundances.

Finally, the region of model parameter $\mathcal{C}$ that 
can fit the BBN observation results well is investigated.
For this purpose, interpolation is performed for 
the light elements abundances at $\mathcal{C}=0.01$ ${\rm s^2}$ and 0.
To constrain this free parameter $\mathcal{C}$, 
two most precisely measured abundances, $\rm Y_p$ and D/H, are used, and 
one inequality condition that positive $\mathcal{C}$ should satisfy is extracted.
Once again, it is confirmed  that the case of D/H can give a much stronger constraint than the case of $\rm Y_p$. 
The negative region of $\mathcal{C}$ has already been excluded 
because there is no appropriate numerical solution for BBN.
In addition, $\mathcal{A} = 1/(6 M^2)$ should be positive because the constant $M$ could be physically interpreted as mass when $\mathcal{B} = L_m = 0$ under some theoretical conditions (e.g. the conformal symmetry) \cite{92Mukhanov, 05Mukhanov}.
As a result, the range $\mathcal{C}$
allowed through the BBN observations in the eSM is
$0 < \mathcal{C} < 1.3 \times 10^{-5}\, {\rm s^2}$.


In conclusion, the eSM can have the standard, and non-standard cosmic evolutions 
due to the higher-order effects in the radiative-dominated universe. 
There exists a parameter space of $\mathcal{C}$ that can satisfy both 
the BBN constraints and numerical stabilities 
in the \textit{\textbf{oscillating non-standard cosmic evolution}}.
Currently very precisely measured $\rm Y_p$ and D/H abundances 
have already been very effectively constraining the model free parameter $\mathcal{C}$.
In particular, measurements of these two elements, which will become more accurate in the near future, 
will help to validate the eSM or to distinguish it from the standard and other models 
by further constraining the parameter space.
Ultimately, they will be essential to understanding and 
identifying exactly how the universe evolved in the early days of radiation dominance.

\appendix
\section{Exponentially increasing scale factor} 
\begin{figure}
\centering
\subfigure[]
{\includegraphics[width=0.49\textwidth]{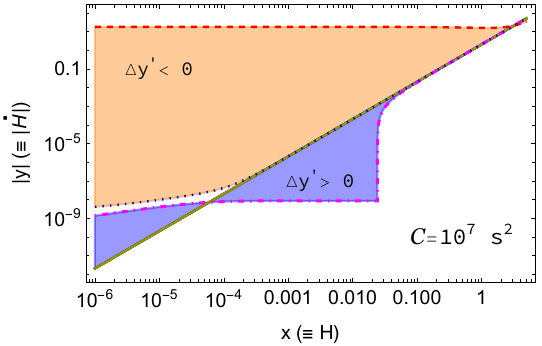}}
\subfigure[]
{\includegraphics[width=0.49\textwidth]{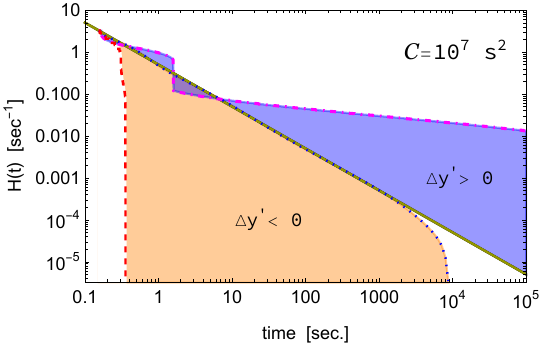}}
\subfigure[]
{\includegraphics[width=0.49\textwidth]{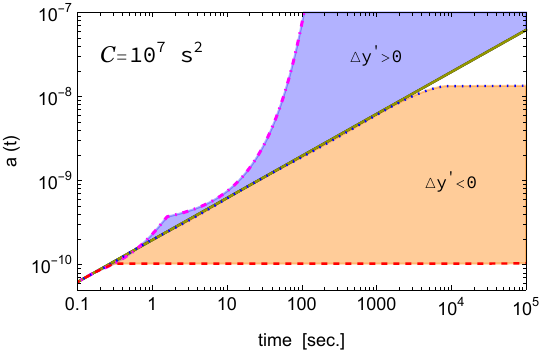}}
\caption{Spaces of the allowed solutions that satisfy the nonlinear cosmic evolution 
equation mentioned in the text. In the phase space (upper left panel, (a)), 
$x$ is the cosmic expansion 
rate $(\equiv H)$ and $|y|$ is the absolute value of the time derivative of 
the cosmic expansion rate $|\dot{H}|$.
The upper right panel (b) shows the allowed cosmic expansion rate $H$ behaviours 
as a function of time. 
Similarly, the bottom panel (c) shows the allowed scale factor $a(t)$ behaviours 
as a function of time.
Here, each orange (blue) region represents the region corresponding to 
$\Delta y^\prime < 0\ (\Delta y^\prime > 0)$ in each panel.
For example, the blue region in the upper left panel shows the correspondence 
between the slowly decreasing cosmic expansion rate (b) and the exponentially 
increasing scale factor (c) at a later point in time compared to standard 
cosmological evolution (black diagonal straight line).
\label{fig:MG_C10to7}}
\end{figure}

This appendix presents an example of the eSM with an exponentially increasing scale factor, 
mentioned but not shown in the main text.
In particular, for this purpose, $\mathcal{C}=10^7 \, {\rm s^2}$, 
which is the case with the greatest effects of the high-order terms, is selected.
In some sense, the MG terms greatly exceed the Einstein term 
representing the standard cosmology, indicating the MG-dominant universe.
Therefore, the previously allowed areas of the non-standard cosmic evolutions 
shown in the main text have been greatly expanded, 
and they have been represented by the blue areas in Fig.\,\ref{fig:MG_C10to7}.
In particular, note that the blue areas, which did not appear in the 
Figs.\,\ref{figNSEC:Numerical_Solutions_phase_space} and 
\ref{figNSEC:Numerical_Solutions}, appear together with the red areas.
The upper left panel in Fig.\,\ref{fig:MG_C10to7} (a) shows the allowed solutions in phase space, 
where $x$ is the cosmic expansion 
rate $H$ and $|y|$ is the absolute value of the time derivative of 
the cosmic expansion rate $|\dot{H}|$.
In the same figure, the upper right and bottom panels (b and c) show the allowed cosmic expansion rate $H$ behaviours 
as a function of time, and the allowed scale factor $a(t)$ behaviours as a function of time, respectively.
Here, the same colors correspond to each other in each panel, and 
the orange (blue) areas correspond to the cases of $\Delta y'<0\,(\Delta y'>0)$.
When compared based on standard cosmic evolution (black diagonal solid line), 
each subpanel (b) and (c) shows 
the case of a slowly decreasing cosmic expansion rate (blue region in (b)) 
and an exponentially increasing scale factor (blue region in (c)), respectively.

\section{Time-temperature relation} 
\begin{figure}
\centering
\includegraphics[width=0.7\textwidth]{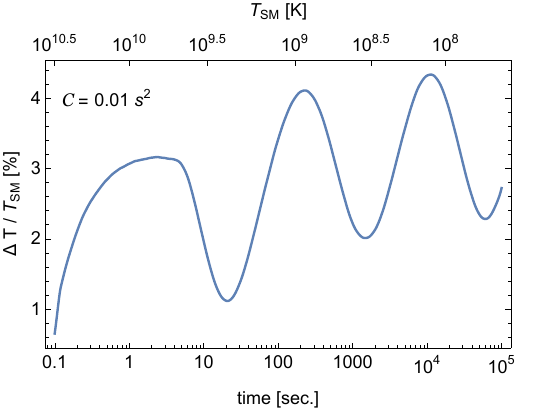}
\caption{Temperature difference between the eSM model with $\mathcal{C}=0.01\, {\rm s^2}$ and 
the SBBN model over time. The difference $\Delta T$ is defined as 
$T_{\rm eSM} - T_{\rm SBBN}$ where $T_{\rm eSM}$ $(T_{\rm SBBN})$ is the temperature in 
eSM (SBBN) model.
At the top of the frame, $T_{\rm SBBN}$ in the SBBN model corresponding to a given time, 
is also indicated in Kelvin (K), 
through which the $T_{\rm eSM}$ in eSM model can be estimated.
Note that this result is closely related to that of Fig.\,\ref{fig:Ratios} (b).
\label{fig:t-T_rel}
}
\end{figure}
In general, comparing two different space-times is not obvious.
Especially in Fig.\,\ref{fig:LightYields} of the main text (where the x-axis is time), 
it is not trivial to compare and understand the temperature of each model.
For this purpose, we present the temperature difference between the eSM model 
with $\mathcal{C}=0.01\, {\rm s^2}$ and the SBBN model over time in Fig.\,\ref{fig:t-T_rel}, 
where the difference $\Delta T$ is defined as $T_{\rm eSM} - T_{\rm SBBN}$.
At the top of the frame, the temperature ($T_{\rm SBBN}$) in the SBBN model corresponding 
to a given time, is indicated in Kelvin (K), 
through which the temperature ($T_{\rm eSM}$) of the eSM model can be estimated.
As a result, 
the eSM model can get hotter by up to $4.4\%$ approximately compared to the SBBN.
The consistency of the results can be checked through Fig.\,\ref{fig:Ratios} (b). 
That is, while the scale factor decreases by several percent and vibrates, 
the temperature increases by several percent and vibrates.
In addition, we can see that this difference naturally disappears 
when we return to the early pre-BBN universe.

\section{Divergence-free property of the tensors from the generalized gravity}
In this Appendix, we briefly show that the covariant divergence of 
the tensors from $\mathcal{A}$ and $\mathcal{B}$ gravity 
vanish independently, 
$T^{\mathcal{A}a}_{\phantom{\mathcal{A}a}b;a} = 0 = T^{\mathcal{B}a}_{\phantom{\mathcal{B}a}b;a}$ by virtue of the Bianchi identities, the algebraic properties of the Riemann curvature tensor, and the commutation relations of covariant derivatives.
To begin with, we check the divergence-free property of the tensor Eq.\,(\ref{EM tensor A}) from 
$\mathcal{A}$ gravity,
\bea 
 && { {8\pi G}  \over \mathcal{A} } T^{\mathcal{A}a}_{\phantom{\mathcal{A}a}b;a}
 =  \Big( {1 \over 2}  R^{2} \delta^{a}_{b} - 2R R^{a}_{b} - 2 \delta^{a}_{b} \Box R + 2 R^{;a}_{\phantom{;a}b}   \Big) _{;a} 
\nonumber
\\ && 
= R R_{;b} - 2 \big( R_{;a} R^a_b + R  R^{a}_{\phantom{a}b;a} \big) - 2 \delta^{a}_{b} R^{;c}_{\phantom{;c}ca} 
+  2\big( R^{;a}_{\phantom{;a}ab} - R^{;c} R^{a}_{\phantom{a}cba}     \big)
\nonumber \\ &&
= R \big( R_{;b} - 2 R^{a}_{\phantom{a}b;a}  \big)  = 0 .
\eea
Here, we use the Bianchi identity $ R^b_{a;b} = {1 \over 2} R_{;a} $ \cite{Hawking and Ellis, 72Weinberg}
and the commutation relation for covariant derivatives of a contravariant vector \cite{ellis etal 12}
$w^{d}_{\phantom{d};bc}   - w^{d}_{\phantom{d};cb}
=  - R^{d}_{\phantom{d}abc} w^a$
.

Next, we 
parenthesize the terms in 
$T^{\mathcal{B}a}_{\phantom{\mathcal{B}a}b;a}$, 
i.e. the divergence of Eq.\,(\ref{EM tensor B}) from $\mathcal{B}$ gravity,
in order to simplify a tedious calculation for checking the vanishing property:
\bea 
&& { {8\pi G}  \over \mathcal{B} } T^{\mathcal{B}a}_{\phantom{\mathcal{B}a}b;a}
= \Big( {1 \over 2 } R^{cd} R_{cd} \delta^{a}_{b}       - \delta^{a}_{b} {R^{cd}}_{;cd}  
		      +   {R}^{ca}_{ \phantom{ca};bc} +       {R}^{c\phantom{b;}a}_{\phantom{a}b;\phantom{a}c} 
- g^{cd} R^{a}_{b;cd}      - 2  R^{ac} R_{bc}  \Big)_{;a}
\nonumber \\ &&
= \Big[ \Big( {R}^{c\phantom{b;}a}_{\phantom{a}b;\phantom{a}ca}    - g^{cd} R^{a}_{b;cda}    \Big)
+ \big(  - {R^{cd}}_{;cdb}  +  {R}^{ca}_{ \phantom{ca};bca}   \big) \Big]
+ \Big[ {1 \over 2 } \big( R^{cd} R_{cd} \big)_{;b}       - 2 \big(   R^{ac} R_{bc}  \big)_{;a}   \Big]
\nonumber \\ &&
= \Big[  \Big(  R^{cd}_{\phantom{cd};a} R^{a}_{\phantom{a}dbc} \Big)   
           +\big(  R_{ab} R^{;a}   +R_{cb;a} R^{ca}  
                      -  R^{a}_{\phantom{a}dbc;a} R^{cd}      - R^{a}_{\phantom{a}dbc} R^{cd}_{\phantom{cd};a}      \big) \Big]
\nonumber \\ &&
 \quad
 + \Big[            - R^{;c} R_{bc} + R^{cd} R_{cd;b} - 2 R^{ac} R_{bc;a}       \Big]
\nonumber \\ &&
= \Big[R_{ab} R^{;a}     - R^{cd} \big( R_{dc;b} - 2 R_{db;c}  \big)     \Big] 
   + \Big[            - R^{;c} R_{bc} + R^{cd} R_{cd;b} - 2 R^{ac} R_{bc;a}       \Big]
= 0.
\eea
In the above calculations, we rely on the mathematical relations \cite{Hawking and Ellis, 72Weinberg} such as properties of the Riemann tensor
\bea 
&&
R_{abcd} = -R_{bacd} = - R_{abdc}    = R_{cdab} ,
\qquad   R^{a}_{\phantom{a}bcd} + R^{a}_{\phantom{a}dbc} +R^{a}_{\phantom{a}cdb} = 0,
 \\ &&
R^{a}_{\phantom{a}bcd;e} + R^{a}_{\phantom{a}bec;d} + R^{a}_{\phantom{a}bde;c} = 0,
\qquad   R^{a}_{\phantom{a}bcd;a}  = R_{bd;c} - R_{bc;d},
\eea
and the generalized commutation relation for any tensor
 \bea 
&& T^{a_{1} \ldots a_{r}}_{\phantom{a_{1} \ldots a_{r}} b_{1} \ldots b_{s};cd}
- T^{a_{1} \ldots a_{r}}_{\phantom{a_{1} \ldots a_{r}} b_{1} \ldots b_{s};dc}
\nonumber \\ &&
= - R^{a_{1}}_{\phantom{a_1}ecd}    T^{e a_{2} \ldots a_{r}}_{\phantom{e a_{2} \ldots a_{r}} b_{1} \ldots b_{s}  }
   - \ldots  
  - R^{a_{r}}_{\phantom{a_r}ecd}    T^{a_{1} \ldots a_{r-1}e}_{\phantom{a_{1} \ldots a_{r-1}e} b_{1} \ldots b_{s}  }
\nonumber \\ &&
\quad 
   +  R^{f}_{\phantom{f}b_{1}cd}    T^{a_{1} \ldots a_{r}}_{\phantom{a_{1} \ldots a_{r}}  f b_{2} \ldots b_{s}  }
    + \ldots
  +  R^{f}_{\phantom{f}b_{s}cd}    T^{a_{1} \ldots a_{r}}_{\phantom{a_{1} \ldots a_{r}}  b_{1} \ldots b_{s-1} f  } .
\eea

\section*{Acknowledgments}
The authors thank Motohiko Kusakabe for valuable discussions on the equation of state and energy 
density of the model.
The authors wish to express their gratitude to Cyril Pitrou for \texttt{PRIMAT}.
This research was supported by a grant from the National Research Foundation of Korea (Grant No. NRF-2020R1A2C3006177, 
Grant No. NRF-2021R1A6A1A03043957, and 2018R1D1A1B07051126).
The work of D.J. was supported by the Institute for Basic Science under IBS-R012-D1.


\begin{thebibliography}{99}
\bibitem{PRD08Weinberg}S. Weinberg, Phys. Rev. D {\bf 77}, 123541 (2008). 
\bibitem{05 H and N} J. Hwang and H. Noh, Phys. Rev. D {\bf 71}, 063536 (2005).
\bibitem{Nojiri:2010wj}
S.~Nojiri and S.~D.~Odintsov,
Phys. Rept. \textbf{505} (2011), 59-144
doi:10.1016/j.physrep.2011.04.001
[arXiv:1011.0544 [gr-qc]].


\bibitem{Nojiri:2017ncd}
S.~Nojiri, S.~D.~Odintsov and V.~K.~Oikonomou,
Phys. Rept. \textbf{692} (2017), 1-104
doi:10.1016/j.physrep.2017.06.001
[arXiv:1705.11098 [gr-qc]].



\bibitem{Starobinsky:1980te}
A.~A.~Starobinsky,
Phys. Lett. B \textbf{91}, 99-102 (1980)
doi:10.1016/0370-2693(80)90670-X

\bibitem{Clifton:2005aj}
T.~Clifton and J.~D.~Barrow,
Phys. Rev. D \textbf{72}, no.10, 103005 (2005)
[erratum: Phys. Rev. D \textbf{90}, no.2, 029902 (2014)]
doi:10.1103/PhysRevD.72.103005
[arXiv:gr-qc/0509059 [gr-qc]].

\bibitem{Lambiase:2006dq}
G.~Lambiase and G.~Scarpetta,
Phys. Rev. D \textbf{74}, 087504 (2006)
doi:10.1103/PhysRevD.74.087504
[arXiv:astro-ph/0610367 [astro-ph]].

\bibitem{Kang:2008zi}
J.~U.~Kang and G.~Panotopoulos,
Phys. Lett. B \textbf{677}, 6-11 (2009)
doi:10.1016/j.physletb.2009.05.006
[arXiv:0806.1493 [astro-ph]].

\bibitem{Kusakabe:2015yaa}
M.~Kusakabe, S.~Koh, K.~S.~Kim and M.-K.~Cheoun,
Phys. Rev. D \textbf{91}, 104023 (2015)
doi:10.1103/PhysRevD.91.104023
[arXiv:1506.08859 [astro-ph.CO]].


\bibitem{Kusakabe:2015ida}
M.~Kusakabe, S.~Koh, K.~S.~Kim and M.-K.~Cheoun,
Phys. Rev. D \textbf{93}, no.4, 043511 (2016)
doi:10.1103/PhysRevD.93.043511
[arXiv:1507.05565 [astro-ph.CO]].


\bibitem{Randall:1999vf}
L.~Randall and R.~Sundrum,
Phys. Rev. Lett. \textbf{83}, 4690-4693 (1999)
doi:10.1103/PhysRevLett.83.4690
[arXiv:hep-th/9906064 [hep-th]].

\bibitem{Randall:1999ee}
L.~Randall and R.~Sundrum,
Phys. Rev. Lett. \textbf{83}, 3370-3373 (1999)
doi:10.1103/PhysRevLett.83.3370
[arXiv:hep-ph/9905221 [hep-ph]].

\bibitem{Binetruy:1999hy}
P.~Binetruy, C.~Deffayet, U.~Ellwanger and D.~Langlois,
Phys. Lett. B \textbf{477}, 285-291 (2000)
doi:10.1016/S0370-2693(00)00204-5
[arXiv:hep-th/9910219 [hep-th]].

\bibitem{Langlois:2000ph}
D.~Langlois,
Phys. Rev. Lett. \textbf{86}, 2212-2215 (2001)
doi:10.1103/PhysRevLett.86.2212
[arXiv:hep-th/0010063 [hep-th]].

\bibitem{Langlois:2000ia}
D.~Langlois,
Phys. Rev. D \textbf{62}, 126012 (2000)
doi:10.1103/PhysRevD.62.126012
[arXiv:hep-th/0005025 [hep-th]].

\bibitem{Sasankan:2016ixg}
N.~Sasankan, M.~R.~Gangopadhyay, G.~J.~Mathews and M.~Kusakabe,
Phys. Rev. D \textbf{95}, no.8, 083516 (2017)
doi:10.1103/PhysRevD.95.083516
[arXiv:1607.06858 [astro-ph.CO]].

\bibitem{Jang:2016rpi}
D.~Jang, M.~Kusakabe and M.-K.~Cheoun,
Phys. Rev. D \textbf{97}, no.4, 043005 (2018)
doi:10.1103/PhysRevD.97.043005
[arXiv:1611.04472 [nucl-th]].

\bibitem{Jang:2018moh}
D.~Jang, Y.~Kwon, K.~Kwak and M.-K.~Cheoun,
Astron. Astrophys. \textbf{650}, A121 (2021)
doi:10.1051/0004-6361/202038478
[arXiv:1812.09472 [astro-ph.CO]].

\bibitem{Kusakabe:2014moa}
M.~Kusakabe, K.~S.~Kim, M.-K.~Cheoun, T.~Kajino, Y.~Kino and G.~J.~Mathews,
Astrophys. J. Suppl. \textbf{214}, 5 (2014)
doi:10.1088/0067-0049/214/1/5
[arXiv:1403.4156 [astro-ph.CO]].


\bibitem{Pitrou:2018cgg}
C.~Pitrou, A.~Coc, J.~P.~Uzan and E.~Vangioni,
Phys. Rept. \textbf{754}, 1-66 (2018)
doi:10.1016/j.physrep.2018.04.005
[arXiv:1801.08023 [astro-ph.CO]].

\bibitem{Pitrou:2020etk}
C.~Pitrou, A.~Coc, J.~P.~Uzan and E.~Vangioni,
Mon. Not. Roy. Astron. Soc. \textbf{502}, no.2, 2474-2481 (2021)
doi:10.1093/mnras/stab135
[arXiv:2011.11320 [astro-ph.CO]].



\bibitem{97 N and H} H. Noh and J. Hwang, Phys. Rev. D {\bf 55}, 5222 (1997).
\bibitem{99 N and H} H. Noh and J. Hwang, Phys. Rev. D {\bf 59}, 047501 (1999).
\bibitem{84 Whitt} B. Whitt, Phys. Lett. {\bf 145B}, 176 (1984).
\bibitem{82 birrell and davis} N. D. Birrell and P. C. W. Davies, {\it Quantum Fields in Curved Space} (Cambridge University Press, England, 1982).
\bibitem{05Mukhanov} V. F. Mukhanov, {\it Physical Foundations of Cosmology} (Cambridge University Press, Cambridge, 2005).
\bibitem{10 Sotiriou and Faraoni} T. P. Sotiriou and V. Faraoni, Rev. Mod. Phys. {\bf 82}, 451 (2010).
\bibitem{Buchbinder} I. L. Buchbinder, S. D. Odintsov and I. L. Shapiro, {\it Effective Action in Quantum Gravity (1st ed.)} (Routledge, New York, 1992) doi:10.1201/9780203758922.
%
\bibitem{08 Capozziello and De Felice} S. Capozziello and A. De Felice, J. Cosmol. Astropart. Phys. {\bf 08}, 10 (2008) 016.
%
\bibitem{Hawking and Ellis} S. W. Hawking and G. F. R. Ellis, {\it The large scale structure of space-time} (Cambridge University Press, New York, 1973).
\bibitem{72Weinberg} S. Weinberg, {\it Gravitation and Cosmology} (Wiley, New York, 1972).
\bibitem{83Barth} N. H. Barth and S. M. Christensen, Phys. Rev. D {\bf 28}, 1876 (1983).

\bibitem{Planck2018} Planck Collaboration, Astron. and Astrophys. {\bf 641}, A10 (2020). 
%
%
%
%
%
%
%
%
%
%
%
%
%
%




%
%


\bibitem{Yun:2022xce}
C.~Yun, J.~Park, M.-K.~Cheoun and D.~Jang,
Phys. Lett. B \textbf{837}, 137652 (2023)
doi:10.1016/j.physletb.2022.137652
[arXiv:2207.09641 [gr-qc]].


%
%
\bibitem{Cooke:2017cwo}
R.~J.~Cooke, M.~Pettini and C.~C.~Steidel,
Astrophys. J. \textbf{855}, no.2, 102 (2018)
doi:10.3847/1538-4357/aaab53
[arXiv:1710.11129 [astro-ph.CO]].
%
%
%
%
\bibitem{ParticleDataGroup:2016lqr}
C.~Patrignani \textit{et al.} [Particle Data Group],
Chin. Phys. C \textbf{40}, no.10, 100001 (2016)
doi:10.1088/1674-1137/40/10/100001


%
\bibitem{Serebrov:2017bzo}
A.~P.~Serebrov, E.~A.~Kolomensky, A.~K.~Fomin, I.~A.~Krasnoshchekova, A.~V.~Vassiljev, D.~M.~Prudnikov, I.~V.~Shoka, A.~V.~Chechkin, M.~E.~Chaikovskiy and V.~E.~Varlamov, \textit{et al.}
Phys. Rev. C \textbf{97}, no.5, 055503 (2018)
doi:10.1103/PhysRevC.97.055503
[arXiv:1712.05663 [nucl-ex]].
%


%
\bibitem{Planck:2015fie}
P.~A.~R.~Ade \textit{et al.} [Planck],
Astron. Astrophys. \textbf{594}, A13 (2016)
doi:10.1051/0004-6361/201525830
[arXiv:1502.01589 [astro-ph.CO]].


%

%
%
\bibitem{Workman:2022ynf}
R.~L.~Workman \textit{et al.} [Particle Data Group],
PTEP \textbf{2022}, 083C01 (2022)
doi:10.1093/ptep/ptac097

%
\bibitem{Aver:2015iza}
E.~Aver, K.~A.~Olive and E.~D.~Skillman,
JCAP \textbf{07}, 011 (2015)
doi:10.1088/1475-7516/2015/07/011
[arXiv:1503.08146 [astro-ph.CO]].
%
%
%
%
\bibitem{Pettini:2012}
M.~Pettini and R.~Cooke, 
Monthly Notices of the Royal Astronomical Society, Volume 425, Issue 4, October 2012, Pages 2477–2486, 
doi.org/10.1111/j.1365-2966.2012.21665.x
[arXiv:1205.3785 [astro-ph.CO]].
%
%
\bibitem{Cooke:2013cba}
R.~Cooke, M.~Pettini, R.~A.~Jorgenson, M.~T.~Murphy and C.~C.~Steidel,
Astrophys. J. \textbf{781}, no.1, 31 (2014)
doi:10.1088/0004-637X/781/1/31
[arXiv:1308.3240 [astro-ph.CO]].
%
\bibitem{Cooke:2016rky}
R.~J.~Cooke, M.~Pettini, K.~M.~Nollett and R.~Jorgenson,
Astrophys. J. \textbf{830}, no.2, 148 (2016)
doi:10.3847/0004-637X/830/2/148
[arXiv:1607.03900 [astro-ph.CO]].
%
\bibitem{Cooke:2017cwo}
R.~J.~Cooke, M.~Pettini and C.~C.~Steidel,
Astrophys. J. \textbf{855}, no.2, 102 (2018)
doi:10.3847/1538-4357/aaab53
[arXiv:1710.11129 [astro-ph.CO]].
%
\bibitem{Balashev:2015hoe}
S.~A.~Balashev, E.~O.~Zavarygin, A.~V.~Ivanchik, K.~N.~Telikova and D.~A.~Varshalovich,
Mon. Not. Roy. Astron. Soc. \textbf{458}, no.2, 2188-2198 (2016)
doi:10.1093/mnras/stw356
[arXiv:1511.01797 [astro-ph.GA]].
%
\bibitem{Riemer-Sorensen:2017pey}
S.~Riemer-S\o{}rensen, S.~Kotu\v{s}, J.~K.~Webb, K.~Ali, V.~Dumont, M.~T.~Murphy and R.~F.~Carswell,
Mon. Not. Roy. Astron. Soc. \textbf{468}, no.3, 3239-3250 (2017)
doi:10.1093/mnras/stx681
[arXiv:1703.06656 [astro-ph.CO]].
%
%
%
%
\bibitem{Vangioni-Flam:2002cvh}
E.~Vangioni-Flam, K.~A.~Olive, B.~D.~Fields and M.~Casse,
Astrophys. J. \textbf{585}, 611-616 (2003)
doi:10.1086/346232
[arXiv:astro-ph/0207583 [astro-ph]].
%
\bibitem{Bania2002}
T.~M.~Bania, R.~T.~Rood and D.~S.~Balser,
Nature. 2002 Jan 3;415(6867):54-7. doi: 10.1038/415054a.
%
%
%
%
\bibitem{Sbordone:2010}
L. Sbordone \textit{et al.},
A$\&$A, 522 (2010) A26, doi:10.1051/0004-6361/200913282

%
%
\bibitem{Cyburt:2008up}
R.~H.~Cyburt and B.~Davids,
Phys. Rev. C \textbf{78}, 064614 (2008)
doi:10.1103/PhysRevC.78.064614
[arXiv:0809.3240 [nucl-ex]].

\bibitem{Cyburt:2008et}
R.~H.~Cyburt, B.~D.~Fields and K.~A.~Olive,
J. Cosmol. Astropart. Phys. 11, 012 
doi.org/10.1088/1475-7516/2008/11/012
%
%
%
\bibitem{Arapoglu:2010rz}
A.~S.~Arapo\u{g}lu, C.~Deliduman and K.~Y.~Ek\c{s}i,
JCAP \textbf{07}, 020 (2011)
doi:10.1088/1475-7516/2011/07/020
[arXiv:1003.3179 [gr-qc]].
%
%

\bibitem{92Mukhanov} V. F. Mukhanov, H. A. Feldmann, and R. H. Brandenberger, Phys. Rep. {\bf 55}, 203 (1992).
%
\bibitem{ellis etal 12} G. F. R. Ellis, R. Maartens and M. A. H. MacCallum, {\it Relativistic Cosmology} (Cambridge University Press, New York, 2012).



\end{thebibliography}
\end{document}